\RequirePackage{amsmath}
\documentclass[final,3p,times,11pt]{elsarticle}
\usepackage[latin1,utf8]{inputenc}
\usepackage[USenglish]{babel}
\usepackage{lmodern}
\usepackage{amssymb}
\usepackage{mathtools}
\usepackage{graphicx}
\usepackage[usenames,dvipsnames]{xcolor}
\usepackage{tikz}
\usepackage{hyperref}
\usepackage{upgreek}
\usepackage{glossaries}
\usepackage{bm}
\usepackage{algorithm}
\usepackage[noend]{algpseudocode}
\usepackage[figurename=Fig.,labelfont=bf,labelsep=quad]{caption}
\usepackage[caption=false]{subfig}
\usepackage{alphalph}
\usepackage{multirow}
\usepackage{chngcntr}
\usepackage[symbol*]{footmisc}
\usetikzlibrary{decorations.text,shapes,arrows,shapes.multipart,calc,positioning,matrix,backgrounds,folding,
 decorations.pathmorphing,spy,fit,decorations.fractals,decorations.markings,external}
\PassOptionsToPackage{hyphens}{url}\usepackage{hyperref}

\hypersetup{
        pdfborder={0 0 0},
        colorlinks=true,
        linkcolor=blue,
        citecolor=blue,
        urlcolor=black,
        pdfmenubar=true,
        pdfnewwindow = true,
        pdfencoding  = unicode,
        pdfauthor={Juan F. Restrepo},
        pdftitle={Automatic estimation of attractor invariants}
}

\glsdisablehyper
\newglossary[algh]{hidden}{acrh}{acnh}{Hidden Acronyms}
\newglossary[slg]{symbols}{sot}{stn}{Lista de Símbolos}
\newacronym{gci}{GCI}{Gaussian kernel correlation integral}
\newacronym{ncii}{NCI}{noise assisted correlation integral}
\newacronym{uci}{UCI}{U-correlation integral}
\newacronym{uci2}{\null$\protect\bm{\protect\inlineUCI}$}{U correlation integral with $\protect\beta=2$}
\newacronym{ucim}{\null$\protect\bm{\protect\inlineUCIm}$}{U correlation integral with  $\protect\beta=m$}
\newacronym{cd}{\ensuremath{{D}}}{correlation dimension}
\newacronym{k2}{\ensuremath{{K_2}}}{correlation entropy}
\newacronym{nl}{\ensuremath{{\sigma}}}{noise level}
\newacronym[longplural=coarse-grained estimators,  shortplural=CgEs, firstplural=Coarse-grained estimators (CgEs)
]{cge}{CgE}{Coarse-grained estimator}
\newacronym{snr}{SNR}{signal to noise ratio}
\newacronym{etal}{et al.}{and others}
\newacronym{pdf}{pdf}{probability density function}
\newacronym{eeg}{EEG}{electroencefalograma}
\newacronym{hrv}{VFC}{variabilidad de frecuencia cardíaca}

\newglossaryentry{sot:genci}
{   type=symbols,
    name={\null$\protect{{G}_{m}\left(h\right)}$},
    description={Integral de correlación generalizada},
    sort={integraldecorrelacióngeneralizada}
}
\newglossaryentry{sot:sci}
{   type=symbols,
    name={\null$\protect{{C}_{m}\left(h\right)}$},
    description={Integral de correlación estándar},
    sort={integraldecorrelaciónestandar}
}
\newglossaryentry{sot:gci}
{   type=symbols,
    name={\null$\protect{{T}_{m}\left(h\right)}$},
    description={Integral de correlación de núcleo Gaussiano},
    sort={integraldecorrelacióndenúcleog}
}
\newglossaryentry{sot:nci}
{   type=symbols,
    name={\null$\protect{S_{m}\left(h\right)}$},
    description={Integral de correlación asistida por ruido},
    sort={Integraldecorrelacionasistida}
}
\newglossaryentry{sot:sumnsci}
{   type=symbols,
    name={\null$\protect{\widehat{C}_{m}\left(h\right)}$},
    description={Suma de correlación estándar},
    sort={sumadecorrelacionestandar}
}
\newglossaryentry{sot:uci}
{   type=symbols,
    name={\null$\protect\UCi{m}{\protect\beta}\protect\E{h}$},
    description={Integral de correlación U},
    sort={Integraldecorrelacionu}
}
\newglossaryentry{sot:uci2}
{   type=symbols,
    name={\null$\protect\UCi{m}{\protect\beta=2}\protect\E{h}$},
    description={Integral de correlación U con $\protect\beta=2$},
    sort={Integraldecorrelacionubdos}
}
\newglossaryentry{sot:ucim}
{   type=symbols,
    name={\null$\protect\UCi{m}{\protect\beta=m}\protect\E{h}$},
    description={Integral de correlación U con $\protect\beta=m$},
    sort={Integraldecorrelacionubm}
}
\newglossaryentry{sot:cdu}
{   type=symbols,
    name={\null$\protect\CDu\protect\E{h}$},
    description={Estimador dependiente de la escala para la dimensión de correlación basado en la
    $\protect\UCi{m}{\beta=m}\protect\E{h}$},
    sort={estimadordeescalaparadu}
}
\newglossaryentry{sot:cku}
{   type=symbols,
    name={\null$\protect\CKu\protect\E{h}$},
    description={Estimador dependiente de la escala para la dimensión de correlación basado en la 
    $\protect\UCi{m}{\beta=m}\protect\E{h}$},
    sort={estimadordeescalaparaku}
}
\newglossaryentry{sot:csu}
{   type=symbols,
    name={\null$\protect\CSu\protect\E{h}$},
    description={Estimador dependiente de la escala para la dimensión de correlación basado en la 
    $\protect\UCi{m}{\beta=m}\protect\E{h}$},
    sort={estimadordeescalaparasu}
}
\newglossaryentry{sot:cdn}
{   type=symbols,
    name={\null$\protect\CDn\protect\E{h}$},
    description={Estimador dependiente de la escala para la dimensión de correlación basado en la 
    $\protect T_{m}\protect\E{h}$},
    sort={estimadordeescalaparadn}
}
\newglossaryentry{sot:ckn}
{   type=symbols,
    name={\null$\protect\CKn\protect\E{h}$},
    description={Estimador dependiente de la escala para la entropía de correlación basado en la 
    $\protect T_{m}\E{h}$},
    sort={estimadordeescalaparakn}
}
\newglossaryentry{sot:csn}
{   type=symbols,
    name={\null$\protect\CSn\protect\E{h}$},
    description={Estimador dependiente de la escala para el nivel de ruido basado en la 
    $\protect T_{m}\protect\E{h}$},
    sort={estimadordeescalaparasn}
}
\newglossaryentry{sot:Delu}
{   type=symbols,
    name={\null$\protect{\Delta_{m}^{U}\protect\E{h}}$},
    description={Funcional para el nivel de ruido basado en $\protect\UCi{m}{\protect\beta=m}\protect\E{h}$},
    sort={funcionalnivelderuido}
}
\newglossaryentry{sot:heavi}
{   type=symbols,
    name={\null$\protect{H\left(t \right)}$},
    text={función escalón},
    description={Función escalón},
    sort={funcionescalon}
}
\newglossaryentry{sot:igamma}
{   type=symbols,
    name={\null$\protect\Igf{a}{t}$},
    description={Función Gamma incompleta superior},
    sort={funcióngammain}
}
\newglossaryentry{sot:gamma}
{   type=symbols,
    name={\null$\protect\Gf{a}$},
    description={Función Gamma},
    sort={funcióngamma}
}
\newglossaryentry{sot:1f1}
{   type=symbols,
    name={\null$\protect\Hf{a}{b}{t}$},
    description={Función Hipergeométrica Confluente},
    sort={funcionhipergeometricaconfluente}
}
\newglossaryentry{sot:2f1}
{   type=symbols,
    name={\null$\protect\HF{a}{b}{c}{t}$},
    description={Función Hipergeométrica de Gauss},
    symbol={\null$\protect\HF{a}{b}{c}{t}$},
    sort={funcionhipergeometricadegaus}
}
\newglossaryentry{sot:pdf}
{   type=symbols,
    name={\null$\protect{f_{t}}$},
    description={Función de densidad de probabilidad de la variable aleatoria \null$t$},
    sort={funciondedensidad}
}
\newglossaryentry{sot:cdf}
{   type=symbols,
    name={\null$\protect{F_{t}}$},
    description={Función de distribución acumulada de la variable aleatoria \null$t$},
    sort={funciondedistribucionacumulada}
}
\newglossaryentry{sot:ccdf}
{   type=symbols,
    name={\null$\protect{\widetilde{F_{t}}}$},
    description={Función de distribución acumulada complementaria de la variable aleatoria \null$t$},
    sort={funciondedistribucionacumulada}
}
\newglossaryentry{sot:deltaD}
{   type=symbols,
    name={\null$\protect{\delta_{X_{i}}}$},
    description={Función delta de Dirac $\protect{\delta\protect\E{x-X_{i}}}$},
    sort={funciondeltad}
}
\newglossaryentry{sot:delta}
{   type=symbols,
    name={\null$\protect{\delta\left[n\right]}$},
    description={Distribución delta de Kroenecker},
    sort={distribuciondelta}
}
\newglossaryentry{sot:mnorm}
{   type=symbols,
    name={\null$\protect\mathcal{N}\protect\E{\protect\bm{0},\protect\bm{\protect\Sigma^2}}$},
    description={Distribución normal multivariada de media cero y matriz de covarianza $\protect\bm{\protect\Sigma^2}$},
    sort={distribucionnormalmulti}
}
\newglossaryentry{sot:norm}
{   type=symbols,
    name={\null$\protect\mathcal{N}\protect\E{0,\protect\sigma^2}$},
    description={Distribución normal de media cero y varianza $\protect\sigma^2$},
    sort={distribucionnormal}
}
\newglossaryentry{sot:unif}
{   type=symbols,
    name={\null$\protect\mathcal{U}\protect\E{a,b}$},
    description={Distribución uniforme con soporte en $\protect\E{a,b}$},
    sort={distribucionuniforme}
}
\newglossaryentry{sot:chi}
{   type=symbols,
    name={\null$\protect\inlineChi$},
    description={Distribución Chi},
    symbol={\null$\protect\inlineChi$},
    sort={distribucionchi}
}
\newglossaryentry{sot:chis}
{   type=symbols,
    name={\null$\protect\inlineChiS$},
    description={Distribución Chi cuadrada},
    symbol={\null$\protect\inlineChiS$},
    sort={distribucionchicuadrada}
}
\newglossaryentry{sot:l}
{   type=symbols,
    name={\null$L$},
    description={Número de vectores de estado},
    sort={numerodevectoresdeestado}
}
\newglossaryentry{sot:d0}
{   type=symbols,
    name={\null$d_{0}$},
    description={Dimensión de conteo por cajas},
    sort={dimensiondeconteoporcajas}
}
\newglossaryentry{sot:m0}
{   type=symbols,
    name={\null$m_{0}$},
    description={Mínima dimensión de inmersión},
    sort={minimadimensiondeinmersion}
}
\newglossaryentry{sot:m}
{   type=symbols,
    name={\null$m$},
    description={Dimensión de inmersión},
    sort={dimensiondeinmersion}
}
\newglossaryentry{sot:beta}
{   type=symbols,
    name={\null$\beta$},
    description={Dimensión del ruido utilizado en el algoritmo de correlación asistido por ruido - Grados de libertad},
    sort={dimensiondelruido}
}
\newglossaryentry{sot:tau}
{   type=symbols,
    name={\null$\protect{\tau}$},
    description={Retardo de inmersión},
    sort={retardodeinmesion}
}
\newglossaryentry{sot:h}
{   type=symbols,
    name={\null$\protect{h}$},
    description={Escala o umbral},
    sort={escala}
}
\newglossaryentry{sot:n}
{   type=symbols,
    name={\null$\protect{N}$},
    description={Longitud de la serie temporal},
    sort={longituddelaserietemp}
}
\newglossaryentry{sot:Q}
{   type=symbols,
    name={\null$\protect{Q}$},
    description={Número de comparadores que  componen el algoritmo de correlación - Número  de distancias entre pares de
    vectores de estado},
    sort={Numerodecomparadores}
}
\newglossaryentry{sot:mu}
{   type=symbols,
    name={\null$\protect{\mu_{\omega}}$},
    description={Realización de ruido utilizada en el algoritmo de correlación asistido por ruido},
    sort={realizaciónderuido }
}
\newglossaryentry{sot:distinf}
{   type=symbols,
    name={\null$\protect\norm{\cdot}_{\protect\infty}$},
    description={Distancia Infinito},
    sort={distanciainfinito}
}
\newglossaryentry{sot:dist}
{   type=symbols,
    name={\null$\protect\norm{\cdot}$},
    description={Distancia Euclídea},
    sort={distanciaeuclidea}
}
\newglossaryentry{sot:sdist}
{   type=symbols,
    name={\null$\protect \norm{\cdot}^2$},
    description={Cuadrado de la distancia Euclídea},
    sort={cuadradodeladistanciaeuclidea}
}
\newglossaryentry{sot:Z2}
{   type=symbols,
    name={\null$\protect z_\omega$},
    description={Cuadrado de la distancia Euclídea entre dos vectores de estado},
    sort={cuadradodeladistanciaeuclideaentrevectores}
}
\newglossaryentry{sot:Z}
{   type=symbols,
    name={\null$\protect\Z_{\omega}$},
    description={Distancia Euclídea entre dos vectores de estado},
    sort={distanciaeuclideaentrevectores}
}
\newglossaryentry{sot:varn}
{   type=symbols,
    name={\null$\protect{\sigma_{\protect\eta}}$},
    description={Varianza del ruido $\protect\eta$ presente en la serie temporal},
    sort={varianzadelruido}
}
\newglossaryentry{sot:vars}
{   type=symbols,
    name={\null$\protect{\sigma_{s}}$},
    description={Varianza de la serie temporal s},
    sort={varianzadelaserietemporal}
}
\newglossaryentry{sot:apen}
{   type=symbols,
    name={\null$\protect{Ap\!E\!n\protect\E{m,h,N}}$},
    description={Entropía aproximada},
    sort={entropiaaproximada}
}
\newglossaryentry{sot:apenmax}
{   type=symbols,
    name={\null$\protect{Ap\!E\!n_{m\!a\!x}}$},
    description={Entropía aproximada máxima},
    sort={entropiaaproximadamax}
}
\newglossaryentry{sot:hmax}
{   type=symbols,
    name={\null$\protect{h_{m\!a\!x}}$},
    description={Escala máxima},
    sort={escalaoptima}
}

\setglossarystyle{long}
\setcitestyle{square}

\counterwithout{figure}{section}

\graphicspath{{./figs/}{./figs/tikz/}}
\DeclareGraphicsExtensions{.eps,.pdf}
\makeatletter
\def\ps@pprintTitle{%
 \let\@oddhead\@empty
 \let\@evenhead\@empty
 \def\@oddfoot{}%
 \let\@evenfoot\@oddfoot}
\makeatother
\newcommand{\RR}{\mathbb{R}}      
      
\DeclareRobustCommand{\rbeta}{{\mathpalette\irbeta\relax}}
\newcommand{\irbeta}[2]{\raisebox{\depth}{$#1\beta$}} 
\DeclareRobustCommand{\rchi}{{\mathpalette\irchi\relax}}
\newcommand{\irchi}[2]{\raisebox{\depth}{$#1\chi$}} 
\newcommand{\inlineChi}{\rchi_{{\scriptscriptstyle{\rbeta}}}}
\newcommand{\inlineChiS}{\rchi_{{\scriptscriptstyle{\rbeta}}}^{\scriptscriptstyle{2}}}

\newcommand{\EA}[1]{e^{{{#1}\mathord{\vphantom{#1}\kern-\nulldelimiterspace}}}}
\newcommand{\EB}[2]{e^{{{#2}\mathord{\left/{\vphantom{{#2}{#1}}}\right.\kern-\nulldelimiterspace}{#1}}}}
\newcommand{\Ln}[1]{\ln#1}

\newcommand{\Pdf}[3]{f_{{#1}}^{#2}\!\E{#3}}

\newcommand{\E}[1]{\left(#1\right)}      
\newcommand{\Q}[1]{\left[#1\right]}      
\newcommand{\W}[1]{\left\{#1\right\}}      
\newcommand{\p}{\!+\!}      
\newcommand{\m}{\!-\!}      
\newcommand{\Frac}[1]{\frac{1}{#1}}

\newcommand{\norm}[1]{\left\lVert#1\right\rVert}
\newcommand{\vv}[1]{\bm{#1}}

\newcommand{\Di}[1]{\,\text{d}#1}
\newcommand{\dif}[2]{{\frac{\text{d}#1}{\text{d}#2}}}

\newcommand{\ldif}[2]{{\frac{\text{d}\,{\Ln{#1}}}{\text{d}\,{\Ln{#2}}}}}

\newcommand{\Int}[3]{\ifthenelse{\equal{#1}{\infty}}{\int_{-#1}^{#1}{#3}}{\int_{#1}^{#2}{#3}}}
\newcommand{\Igf}[2]{\Gamma\E{#1,\,#2}}

\newcommand{\Gf}[1]{\Gamma\E{#1}}

\newcommand{\Hf}[3]{\,\,\!_{\!_{1}}\!F_{\!_{1}}\!\!\E{#1,#2;#3}}
\newcommand{\HF}[4]{F\E{#1,#2;#3;#4}}

\newcommand{\UCi}[2]{U_{#1}^{#2}\!}
\newcommand{\CDu}{D_{m}^{\scriptscriptstyle{U}}}
\newcommand{\CDn}{D_{m}^{\scriptscriptstyle{T}}}
\newcommand{\CKu}{K_{m}^{\scriptscriptstyle{U}}}
\newcommand{\CKn}{K_{m}^{\scriptscriptstyle{T}}}
\newcommand{\Del}{\Delta_{m}^{\scriptscriptstyle{U}}}

\newcommand{\logD}{\dot{D}_{m}^{\beta}}
\newcommand{\logDmb}[2]{\dot{D}^{#1}_{#2}}
\newcommand{\logDm}{\dot{D}_{m}^{\beta=m}}
\newcommand{\CSu}{\sigma_{m}^{\scriptscriptstyle{U}}}
\newcommand{\CSn}{\sigma^{\scriptscriptstyle{T}}_{m}}
\newcommand{\itau}{\Delta t \tau}
\DeclareRobustCommand{\rUCI}{{\mathpalette\irUCI\relax}}
\newcommand{\irUCI}[2]{\raisebox{\depth}{$#1\text{UCI}$}} 
\makeatletter
\newcommand{\raisemath}[1]{\mathpalette{\raisem@th{#1}}}
\newcommand{\raisem@th}[3]{\raisebox{#1}{$#2#3$}}
\makeatother
\newcommand{\inlineUCI}{\rUCI^{\raisemath{-1pt}{\beta=2}}}
\newcommand{\inlineUCIm}{\rUCI^{\raisemath{-1pt}{\beta=m}}}
\makeatletter
\newcommand*\rel@kern[1]{\kern#1\dimexpr\macc@kerna}
\newcommand*\widebar[1]{%
  \begingroup
  \def\mathaccent##1##2{%
    \rel@kern{0.8}%
    \overline{\rel@kern{-0.8}\macc@nucleus\rel@kern{0.2}}%
    \rel@kern{-0.2}%
  }%
  \macc@depth\@ne
  \let\math@bgroup\@empty \let\math@egroup\macc@set@skewchar
  \mathsurround\z@ \frozen@everymath{\mathgroup\macc@group\relax}%
  \macc@set@skewchar\relax
  \let\mathaccentV\macc@nested@a
  \macc@nested@a\relax111{#1}%
  \endgroup
}
\makeatother
\newcommand{\z}{\tilde{z}}
\newcommand{\Z}{\tilde{z}}

\newcommand{\RVcolor}{dashed~black~line}
\begin{document}
\begin{frontmatter}
\title{Automatic estimation of attractor invariants}
\author[l1,l2]{Juan F. Restrepo\corref{c1}}
\ead{jrestrepo@bioingenieria.edu.ar}
\author[l1,l2]{Gastón Schlotthauer}
\address[l1]{Laboratorio de Señales y Dinámicas no Lineales,  Facultad de Ingeniería,  Universidad Nacional de Entre
Ríos,\\ Ruta Prov. 11 Km. 10. Oro Verde-Entre Ríos, Argentina.}
\address[l2]{Instituto de Investigación y Desarrollo en Bioingeniería y Bioinformática, Argentina.}
\cortext[c1]{Corresponding author. Tel: +54 0343 4975 100 (122)}
\cortext[c2]{This work was  supported  by  the  National  Scientific  and  Technical  Research Council (CONICET) of
Argentina, the National University of Entre Ríos (UNER), CITER and the Grants: PID-$6171$ (UNER), PICT-$2012$-$2954$
(ANPCyT-UNER), and PIO-$14620140100014$CO (CITER-CONICET).}
\cortext[c3]{This is  a post-peer-review,  pre-copyedit version of  an article published in  Nonlinear Dynamics.
The final authenticated version is available online at: http://dx.doi.org/10.1007/s11071-017-3974-3}
\begin{abstract}
The invariants of an  attractor  have  been  the  most  used  resource  to characterize a nonlinear dynamics.  Their
estimation is a challenging endeavor in short-time series and/or in presence of noise.  In this article,  we present
two new coarse-grained estimators for the correlation dimension and for the correlation entropy.  They can be easily
estimated from the calculation of two U-correlation integrals.  We  have also developed an algorithm that is able to
automatically obtain these invariants and the noise level in order to process large data sets.  This method has been
statistically tested through simulations in low-dimensional systems.  The results show that it is robust in presence
of noise and short data lengths.  In comparison with similar approaches, our algorithm outperforms the estimation of
the correlation entropy.
\end{abstract}
\end{frontmatter}
\section{\label{intro}Introduction}
\glsunset{nl}
\Gls{cd} and \gls{k2}  are quantities that characterize the  natural measure of an attractor.  The  first one can be
interpreted as the dimension of the attractor.  It is a  rough measure of the effective number of degrees of freedom
(number of variables) involved in the dynamical process.  The correlation  entropy is a measure of the rate at which
pairs  of  nearby  orbits diverge.  In  other  words,  it  gives us  an  estimate  of  the  unpredictability  of the
system~\cite{Diks1999}.

The correlation dimension  and entropy allow us  to measure the complexity of  a dynamical system.  Typically,  more
complex systems have higher  dimensions and larger values of entropy.  More important,  these  quantities help us to
identify changes in the system's complexity.  For this reason,  they have become very popular and widely used in the
biomedical field~\cite{Strogatz2014},  not only to characterize the  dynamics of physiological systems,  but also to
detect different types of pathologies.  For  example,  Choi~\glsname{etal} used the correlation dimension calculated
from voice signal to characterize the dynamics of  the phonatory system in normal and pathological conditions.  They
reported  a  mean  value  of  $D=1.57$  for  normal  voices  and  an  increased  dimension  value  for  pathological
ones~\cite{Choi2012}.  In Hosseinifard~\glsname{etal}~\cite{Hosseinifard2013} used  \gls{cd},  among other nonlinear
measures,  extracted  from  electroencephalographic  signals to  classify  between  depressed  and  normal patients.
Combining these features with machine learning  techniques,  they achieved a $90\%$ classification accuracy.  On the
other hand,  in~\cite{Gao2012,Gao2015} the authors illustrated how  \gls{k2} and other related entropy measures have
been used to characterize different biological phenomena.

Since the natural measure of an attractor is invariant under the dynamical evolution, both \gls{cd} and \gls{k2} can
be easily estimated from  indirect  time  measurements  of  one  of  the  system variables~\cite{Kantz2004}. For an
observed stationary time series of length $N$,  $\W{x_n}_{n=1}^{N}$, the reconstructed $m$-dimensional delay vectors
$\vv{x_{i}}=\E{x_i,x_{i+\tau}\dots,{x_{i-\E{m-1}\tau}}}$,  $i=1,2,\dots,L=N-\E{m-1}\tau$,  must be formed.  Here $m$
is the embedding  dimension and $\tau$ is the  embedding lag.  As a collection,  these delay  vectors constitute the
reconstructed attractor of the system in $\RR^{m}$.  From another  point of view,  this collection can be thought as
samples drawn from the natural measure of the system.  These samples can be used to estimate the invariants \gls{cd}
and \gls{k2} using the  correlation integral $G_m\E{h}$.  This last quantity is defined  as the probability that the
distance $\z=\norm{\vv{x_{i}}-\vv{x_{j}}}$ between two randomly  selected $m$-dimensional delay vectors $\vv{x_{i}}$
and $\vv{x_{j}}$ is smaller than a value $h$~\cite{Diks1999}:

\begin{equation}
    G_{m}\E{h} = \int{g\E{h,\z}\Pdf{\Z}{m}{\z}\Di{\z}},\label{eq:GeneralCI}
\end{equation}
where $g$  is a  kernel function  and $\Pdf{\Z}{m}{\z}$  is the  \gls{pdf} of  the distance  between pairs  of delay
vectors.  In this article,  the Euclidean  distance is used,  but others measures of  distance can be employed.  The
Grassberger and Procaccia correlation integral \gls{sot:sci} uses as kernel function $g\E{h,\z}=H\E{1-\z/h}$,  where
$H\E{\cdot}$ is the Heaviside step function~\cite{Grassberger1983b}.  On the other hand, the \gls{gci} \gls{sot:gci}
proposed  by  Diks~\glsname{etal}~\cite{Diks1996}  adopts  $g\E{h,\z}=\exp\E{-\z^2/4h^2}$.  For  deterministic  time
series and in absence of noise, both correlation integrals scale as:
\begin{align}
    C_{m}\E{h}=T_{m}\E{h}&\sim h^{D}\EA{-m\itau K_2}. \nonumber\\
                         & \text{for}\quad m\to \infty\,\,\,\text{and}\,\,\,h\to0,\nonumber
\end{align}
where $\Delta t =1/f_{s}$,  being $f_{s}$ the sampling frequency  of the time series.  This behavior is due to that,
in the attractor of a  chaotic dynamical system,  the expected density of points within a  ball of small radius $h$,
scales as $h^D$.  Thus,  the greater the correlation dimension,  the  more complex the attractor,  and the longer it
takes to revisit the same region in it~\cite{Chan2013}.  On the other hand, the $\EA{-m\itau K_2}$ factor is related
to the exponential divergence of nearby trajectories.  The higher the entropy values,  the faster those trajectories
diverge~\cite{Frank1993,Grassberger1983}.

The main  advantage of the  \gls{gci} over the  Grassberger and Procaccia  correlation integral  is that  the former
allows us to model the influence of additive noise on  the scaling law.  For a time series with added white Gaussian
noise of variance $\sigma^2$, the scaling rule is~\cite{Diks1996,Yu2000,Nolte2001}:
\begin{align}
    {T_{m}\E{h}} \sim&\,\,{h^m}\E{h^2+\sigma^2}^{\frac{D-m}{2}}\EA{-m\itau K_{2}}\nonumber\\
                     & \text{for}\quad m\to \infty\,\,\,\text{and}\,\,\sqrt{h^2+\sigma^2}\to0.\label{eq:gciscale}
\end{align}

In presence  of noise,  it is very  important to be able  to quantify  its level  \gls{nl} in  order to  correct the
estimation of  \gls{cd} and \gls{k2}.  Moreover,  an  imprecise estimate of  these invariants  can lead  to mistaken
conclusions about the studied  phenomena.  In this sense,  an estimate of the noise  level must be reported allowing
other researchers to be aware of the limitations of the data.

The main  problem of  the \gls{gci}  is that  it requires  high values  of $m$  to obtain  a convergent  estimate of
\gls{k2}~\cite{Nolte2001,Small2005}.  This is  a major drawback,  since  high values of  $m$ imply much  longer time
series in order to achieve good estimations.

The  cause  of  this  problem  is  related  with  the  kernel  function  of  the  \gls{gci}.   It  can  be  seen  in
Eq.~(\ref{eq:GeneralCI}) that the \gls{pdf} of the interpoint distance depends on the embedding dimension.  However,
the Gaussian kernel function does not change with $m$, and, for this reason, the convergence of the \glsname{gci} to
the \gls{k2} is slowed down~\cite{Restrepo2016}.

As  a   solution   we   have   recently   proposed  the
\gls{uci}~\cite{Restrepo2016}:
\begin{equation}
    U_{m}^{\beta}\E{h} = \int{\frac{\Igf{\beta/2}{z/h^2}}{\Gf{\beta/2}}\hat{f}_{m}\E{\sigma;z}\Di{z}},\label{eq:uci1}
\end{equation}
where   $\Igf{a}{t}$   is  the   upper   incomplete  Gamma   function,   $\Gf{a}$   is   the   Gamma   function  and
$\hat{f}_{m}\E{\sigma;z}$ is the \gls{pdf} of the squared interpoint distance.  There are two important issues about
this correlation integral that deserve further  analysis.  First,  note that the \gls{uci}'s kernel function,  given
by:
\begin{equation*}
    g\E{h,\beta,z}=\frac{\Igf{\beta/2}{z/h^2}}{\Gf{\beta/2}}
\end{equation*}
has a parameter $\beta$ that is used to incorporate information about the embedding dimension.  In other words, this
kernel function is able to  change according to $m$.  Second,  we are now working with  the square of the interpoint
distance, i.e., $z=\z^2$, to reduce the computational cost.

In order  to find the  scaling behavior of the  \gls{uci},  we need an expression  for the \gls{pdf}  of the squared
distances between pairs of noisy delay vectors $\hat{f}_{m}\E{\sigma;z}$~\cite[see~Eq.~(18)]{Oltmans1997}:
\begin{align}
    \hat{f}_{m}\E{\sigma;z}=&\frac{z^{m/2-1}}{2\E{2\sigma}^{m\!-\!D}} \frac{\Gf{D/2}}{\Gf{m/2}}\nonumber\\
                              &\times \Hf{\frac{m\!-\!D}{2}}{\frac{m}{2}}{-\frac{z}{4\sigma^2}},
                            \label{eq:PDF_DIS}
\end{align}
where  $\Hf{a}{b}{t}$  is  the Kummer's  confluent  hypergeometric  function.  This  distribution  arises  under the
assumption that,  in absence of noise, the \gls{pdf} of the pairwise distances between $m$-dimensional delay vectors
(points in the reconstructed  attractor) behaves as $h^{D-1}$ for a certain  range of $h$ values~\cite{Oltmans1997}.
If each of the coordinates of these vectors is perturbed with white Gaussian noise of variance $\sigma^2$,  then the
\gls{pdf} of the perturbed distances will follow Eq.~(\ref{eq:PDF_DIS}).

The scaling law for the \gls{uci} can be obtained by substituting  Eq.~(\ref{eq:PDF_DIS}) in Eq.~(\ref{eq:uci1}) and
integrating over $z$~\cite{Restrepo2016}:

\begin{align}
 \UCi{m}{\beta}\E{h} =&\frac{\widehat{\phi}}{2}\E{2\sigma}^{D}
 \frac{\Gf{D/2}\Gf{\E{\beta\p m}/{2}}}{\Gf{\beta/2}\Gf{m/2\p1}}\EA{-m\itau K_{2}}\nonumber\\
 & \!\!\times\!\!\E{\frac{h^2}{4\sigma^2}}^{m/2}\!\!\!\HF{\frac{\beta\p m}{2}}{\frac{m\m D}{2}}
 {\frac{m\p2}{2}}{-\frac{h^2}{4\sigma^2}},
 \label{eq:UCI}
\end{align}
where $\hat{\phi}$ is a normalization constant and $\HF{a}{b}{c}{t}$ is the Gauss hypergeometric function.

In practical applications,  once the scaling law has been selected and the correlation integral has been calculated,
there are two main  options  to  estimate  \gls{cd},  \gls{k2}  and  \gls{nl}.  The  first  one is to estimate these
invariants through a nonlinear fitting of  the scaling rule~\cite{Diks1996,Yu2000}.  This method is highly dependent
on the  range of scale  selected to fit  the model,  and there is  not a consensus  about the correct  way to choose
it~\cite{Kantz2004}.  The second  option is to use  coarse-grained estimators.  These  are explicit  expressions for
$D$,  $K_2$,   and  $\sigma$  as  functions  of  $m$  and  $h$~\cite{Diks1999}.   For  example,  in~\cite{Nolte2001}
Nolte~\glsname{etal} have  proposed a set  of coarse-grained functions  based on the  \gls{gci}.  The most important
aspect of their approach is that the calculation of  these functions only depends on the estimation of \gls{sot:gci}
for two consecutive values of the embedding  dimension $m$.  Nevertheless,  given that these estimators are based on
the \gls{gci}, the convergence to \gls{k2} needs high values of $m$ and, therefore, large amounts of data.

In~\cite{Restrepo2016} we have proposed a set of coarse-grained estimators for \gls{cd}, \gls{k2} and \gls{nl} based
on the \gls{uci}.  However, these estimators are highly dependent on a precise estimation of \gls{nl}.

In this article we present a new set of coarse-grained estimators for \gls{cd} and \gls{k2}, based on the \gls{uci},
that only needs the  calculation of two U-correlation integrals i.\ e.\ they do not  require the a priori estimation
of any quantity.  Moreover,  we propose a methodology to automatically estimate these invariants and the noise level
of the time series.

In Section~\ref{sec:TEO} we derive the here proposed coarse-grained estimators.  Section~\ref{sec:AIS} is devoted to
describe the algorithm to automatically estimate  the invariants.  In Section~\ref{sec:Dis} we discuss these results
and  give  some  suggestions  for the  practical  of  the  algorithm.  Finally,  the  conclusions  are  presented in
Section~\ref{sec:Con}.
\section{\label{sec:TEO}Theory}

Following a similar approach to one given by Nolte~\glsname{etal}~\cite{Nolte2001}, the here proposed coarse-grained
estimators are based  on a noise level functional.  This  quantity,  closely related to the noise  level of the time
series,  can be  estimated using two  U-correlation integrals.  In this  work,  the noise  level functional  will be
employed to correct the deviation of the coarse-grained estimators from the scaling law in presence of noise.
\subsection{\label{sub:NLF}Noise level functional}

The noise level functional  based on  the \gls{uci} can be defined as~\cite{Restrepo2016}:
\begin{align}
    \Del\E{h}&=\Frac{2}\E{\logDmb{\beta=m}{m+2}\E{h} -\logDm\E{h}}\nonumber\\
             &\approx\frac{4\sigma^2}{h^2+4\sigma^2} \qquad\text{for}\,\,m\gg D,
    \label{eq:delta}
\end{align}
where
\begin{equation*}
    \logDmb{\beta}{m}\E{h}=\ldif{\UCi{m}{\beta}\E{h}}{h}.
\end{equation*}

As it can be observed from Eq.~(\ref{eq:delta}),  $\Del\E{h}$ is a function that decreases monotonically from $1$ to
$0$ on a scale that is proportional to the value of \gls{nl}.  This functional depends on the logarithmic derivative
of two  U-correlation integrals:  $\UCi{m\phantom{+1}}{\beta=m}\E{h}$ and  $\UCi{m+2}{\beta=m}\E{h}$.  Note that the
parameter $\beta$ is set equal to $m$ in both correlation integrals.

In order illustrate the behavior of this noise level functional, we will use the Henon map: 
\begin{equation*}
\begin{cases} x_{n+1}=1-ax_{n}^{2}+y_{n} &\\ y_{n+1}=bx_{n} & \end{cases},
\end{equation*}
where $a=1.4$ and  $b=0.3$.  This map has been widely  used in the literature to  numerically characterize different
methods devoted to estimate attractor's invariants~\cite{Smith1992, Oltmans1997,  Kugiumtzis1997b, Diks1999, Yu2000,
Nolte2001,  Kantz2004,  Small2005,  Jayawardena2008,Coban2012}.  For this map,  and the given parameters values, the
correlation dimension and correlation entropy are \gls{cd}$=1.22$ and \gls{k2}$=0.3$, respectively~\cite{Yu2000}.

\begin{figure}[t]
\begin{center}
  \subfloat[\label{fig:Delu-a}]{\includegraphics[width=0.5\columnwidth,keepaspectratio]{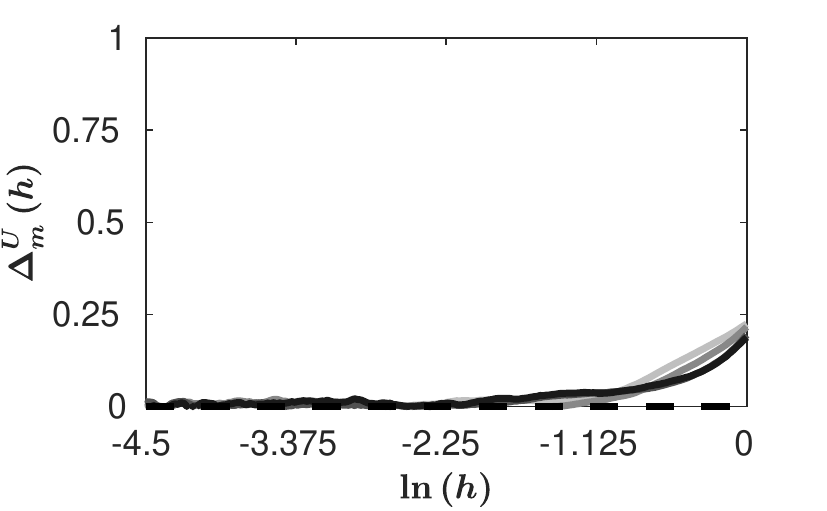}}
  \subfloat[\label{fig:Delu-b}]{\includegraphics[width=0.5\columnwidth,keepaspectratio]{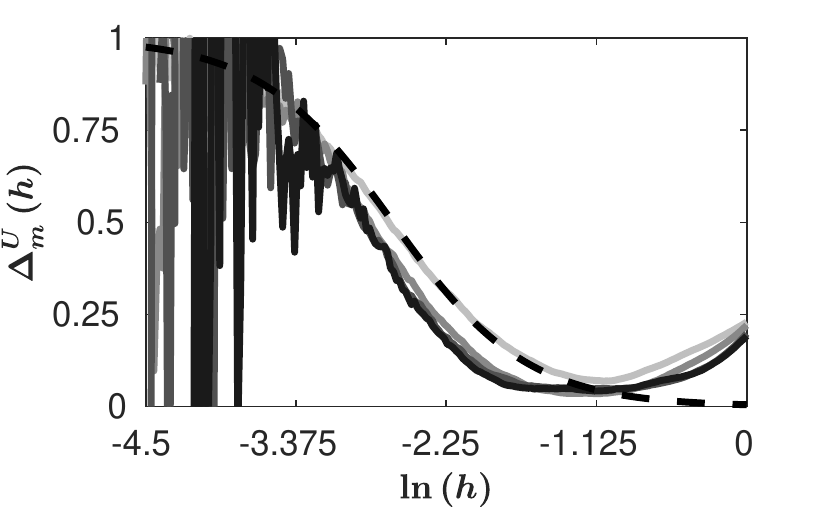}}
\end{center}
\caption{\label{fig:Delu}Henon map:  Noise  level functional  for ${m=\W{2,4,6,8}}$,  the  curves for  different $m$
values  are  color-coded  in  grayscale  where  the  lightest  gray  corresponds  to  $m=2$.   (a)  Noiseless.   (b)
$\protect\sigma=0.05$.  The theoretical curve (right-hand side of Eq.~(\ref{eq:delta})) is shown in \RVcolor.}
\end{figure}

The noise level  functionals for both a  clean and a noisy ($\sigma=0.05$)  realizations of the Henon  map of length
$N=10000$ are shown in Fig.~\ref{fig:Delu}.  Note that in absence of noise (Fig.~\ref{fig:Delu-a}),  this functional
is  close  to  zero  for  a  wide  range of  $h$  values  regardless  the  value  of  $m$.  On  the  other hand,  in
Fig.~\ref{fig:Delu-b} it can be observed that $\Del\E{h}$ falls of  from $1$ to $0$ and the curves for different $m$
values are very close to the theoretical curve for  $\sigma=0.05$.  We must point out that the closest curve of this
functional to  the theoretical  value is  the one corresponding  to $m=2$.  For  larger values  of $m$,  $\Del\E{h}$
slightly deviates from this behavior.
\subsection{\label{ssub:NLE}Noise level coarse-grained estimator}

Starting from Eq.~(\ref{eq:delta}), we have derived the coarse-grained noise level estimator~\cite{Restrepo2016}:

\begin{equation}
 \CSu\E{h}=\frac{h}{2}\sqrt{\frac{\Del\E{h}}{1-\Del\E{h}}}.\label{eq:CoarseS}
\end{equation}
\begin{figure}[t]
\begin{center}
    \subfloat[\label{fig:Su-a}]{\includegraphics[width=0.5\columnwidth,keepaspectratio]{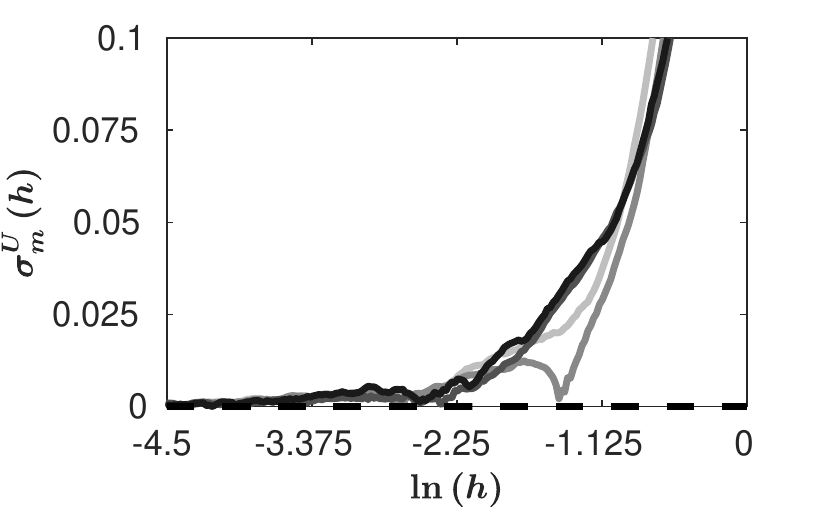}}
    \subfloat[\label{fig:Su-b}]{\includegraphics[width=0.5\columnwidth,keepaspectratio]{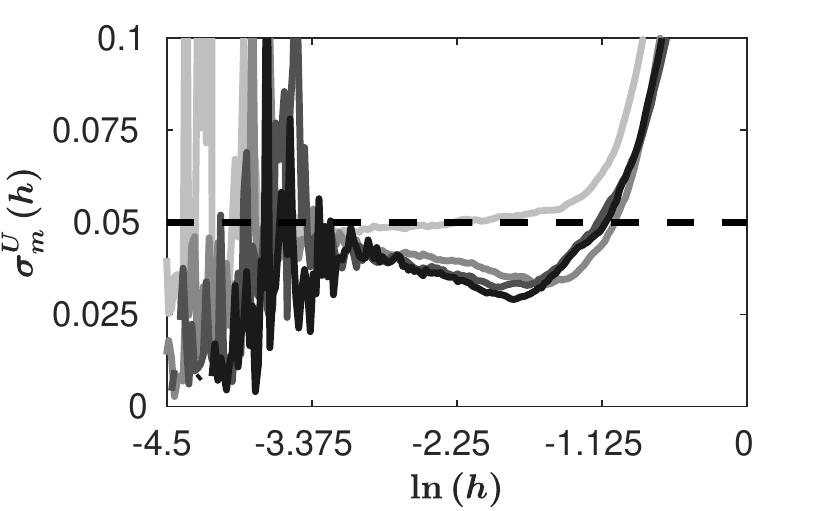}}
\end{center}
\caption{\label{fig:Su}Henon map: Noise level coarse-grained estimator for $m=\W{2,4,6,8}$, the curves for different
$m$  values  are color-coded  in  grayscale where  the  lightest  gray  corresponds  to  $m=2$.  (a) Noiseless.  (b)
$\protect\sigma=0.05$.  The true value of the noise level is shown in \RVcolor.}
\end{figure}

We must  point out that  all time series  used in this  work were rescaled  to have unitary  standard deviation.  In
consequence, $\sigma$ is the noise level after rescaling, i.e.,
\begin{equation*}
 \sigma=\frac{\sigma_n}{\sqrt{\sigma_{c}^{2}+\sigma_{n}^{2}}},
\end{equation*}
where $\sigma_{n}^{2}$ is the noise variance and $\sigma_{c}^{2}$ is the variance of the clean time series.  In this
sense $\sigma\to0$  corresponds to  clean time  series and $\sigma\to1$  implies only  noise.  The \gls{snr}  can be
calculated as:
\begin{equation*}
    \text{SNR}=10\log_{10}\E{\frac{1-\sigma^2}{\sigma^2}}\,\text{dB}.
\end{equation*}

In  Fig.~\ref{fig:Su},  it  is  shown  the  coarse-grained estimator  for  noise  level  as  a  function  of  $h$ for
$m=\W{2,4,6,8,10}$ for both a clean and a noisy  ($\sigma=0.05$) realization of the Henon map with length $N=10000$.
It can be observed in Fig.~\ref{fig:Su-a} that,  for the noiseless case, $\CSu\E{h}$ approaches to zero as $h\to 0$.
On the other hand, Fig.~\ref{fig:Su-b} shows that, when the noise level is increased, $\CSu\E{h}$ underestimates the
value of \gls{nl},  but it  is not too far from its  real value.  Note that the curve of  $\CSu\E{h}$ closest to the
real \gls{nl} is the one calculated for $m=2$ (top curve).  Moreover,  the range of $h$ values where \gls{nl} can be
estimated is larger for $m=2$ than for the other $m$ values.  This suggests that a coarse-grained estimator based on
the \gls{gci} could be better than one based on the \gls{uci} for noise level estimation.
\begin{figure}[t!]
\begin{center}
    \subfloat[\label{fig:Du-a}]{\includegraphics[width=0.5\columnwidth,keepaspectratio]{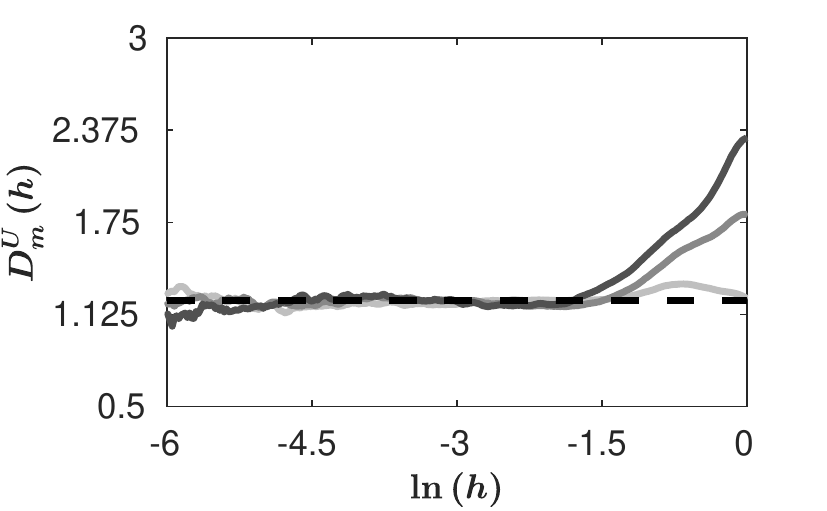}}
    \subfloat[\label{fig:Du-b}]{\includegraphics[width=0.5\columnwidth,keepaspectratio]{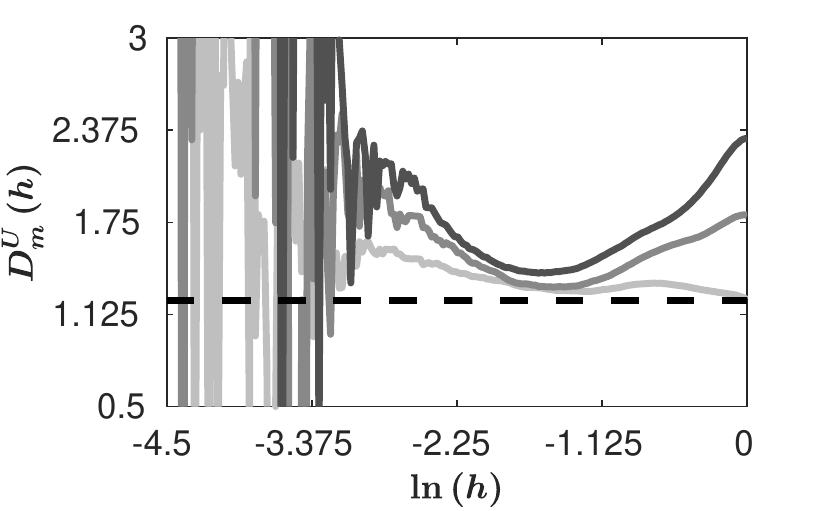}}
\end{center}
\caption{\label{fig:Du}Henon map:  Correlation dimension coarse-grained estimator for $m=\W{4,6,8}$,  the curves for
different $m$ values are color-coded in grayscale where the lightest gray corresponds to $m=4$.  (a) Noiseless.  (b)
$\protect\sigma=0.05$.  The reported value of the correlation dimension is shown in \RVcolor.}
\end{figure}
\subsection{\label{ssub:DE}Correlation dimension coarse-grained estimator}

Here we present a new coarse-grained estimator for \gls{cd} (see Appendix~\ref{sec:A1} for deduction) which is given
by:
\begin{align}
    \CDu\E{h}=&\logDm\E{h}+\frac{\Del\E{h}}{1-\Del\E{h}}\bigg[\logDm\E{h}\nonumber\\
            &\p2\E{m\m1}\E{\frac{\UCi{m}{\beta=m-2}\E{h}}{\UCi{m}{\beta=m}\E{h}}\m1}\bigg].\label{eq:CoarseD}
\end{align}

The  calculation   of  $\CDu\E{h}$  requires   the  estimation  of   $\Del\E{h}$  and  two   correlation  integrals:
$\UCi{m}{\beta=m}\E{h}$ and $\UCi{m}{\beta=m-2}\E{h}$.  In~\cite{Restrepo2016} we  have shown that if $\sigma\ll h$,
then $\logDm\E{h}\to D$.  Moreover,  under the  same condition $\Del\E{h}\to0$.  Replacing in Eq.~(\ref{eq:CoarseD})
we have  that: 
\begin{equation*}
  \CDu\E{h}\to \logDm\E{h}\to  D,  \quad \text{for}\,\,\sigma\ll h. 
\end{equation*}
This  means  that,  in absence  of  noise,  this coarse-grained  estimator  is  the  logarithmic  derivative  of the
\gls{uci},  which tends to  \gls{cd}.  On the other hand,  if noise  is present,  the second term in  the right-hand
side of Eq.~(\ref{eq:CoarseD}) helps to correct the estimation.

Since the  \gls{gci} is a  special case of  the \gls{uci} for  $\beta=2$.  The coarse-grained estimator  of \gls{cd}
proposed   by   Nolte~\glsname{etal}    in~\cite{Nolte2001}   can   be   easily    derived   from   Eq.~(\ref{eq:D5})
(see~Appendix~\ref{sec:A1.1}).

In Fig.~\ref{fig:Du},  the estimations of  $\CDu\E{h}$ for both a clean and a  noisy ($\sigma=0.05$) realizations of
the Henon  map are shown for  $m=\W{4,6,8}$.  In absence of noise,  it  can be seen in  Fig.~\ref{fig:Du-a} that the
estimator $\CDu\E{h}$  oscillates near the reported  value of correlation dimension  ($D=1.22$) for a  wide range of
scales,  regardless  the  value of  $m$.  In  contrast,  Fig.~\ref{fig:Du-b}  shows  that  $\CDu\E{h}$ overestimates
\gls{cd}.  Furthermore,  the best  estimation is reached with  the lowest value of  the embedding dimension,  $m=4$.
This is related with the previously discussed scaling behavior of $\Del\E{h}$.  However, it can be observed that the
values of $\CDu\E{h}$ are close to the reported value of \gls{cd}.
\begin{figure}[t!]
\begin{center}
    \subfloat[\label{fig:Ku-a}]{\includegraphics[width=0.5\columnwidth,keepaspectratio]{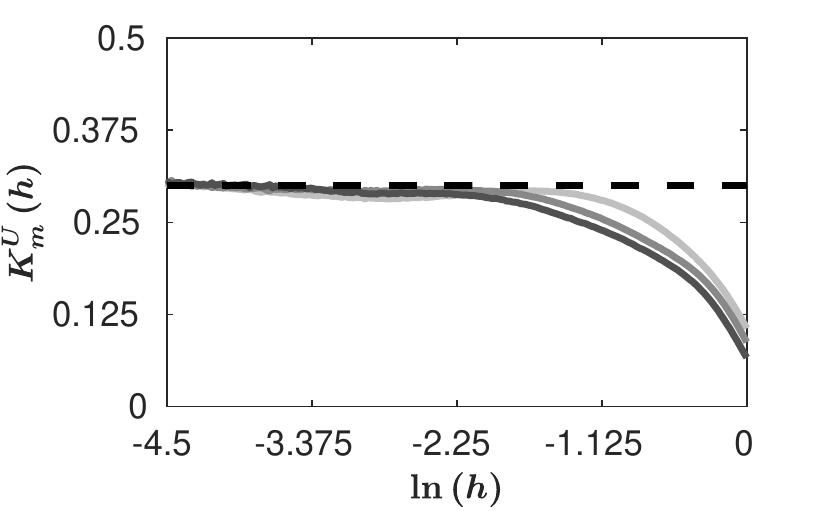}}
    \subfloat[\label{fig:Ku-b}]{\includegraphics[width=0.5\columnwidth,keepaspectratio]{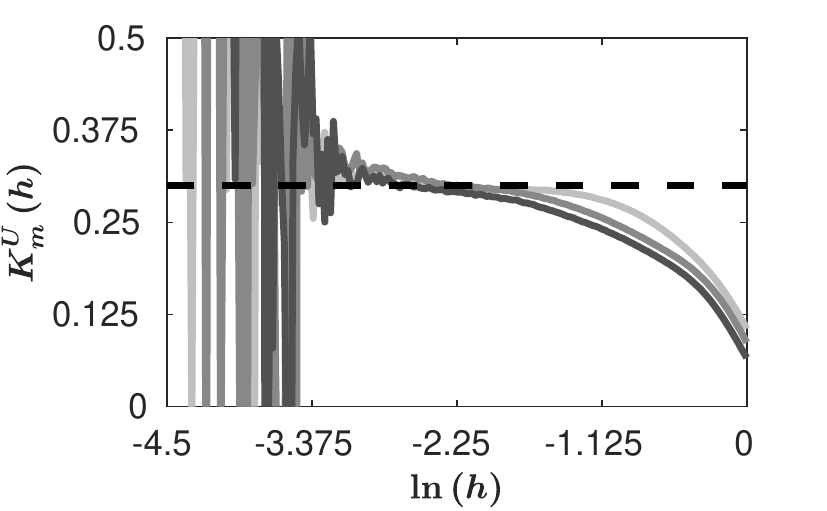}}
\end{center}
\caption{\label{fig:Ku}Correlation entropy coarse-grained estimator for $m=\W{4,6,8}$,  the curves for different $m$
values  are  color-coded  in  grayscale  where  the  lightest  gray  corresponds  to  $m=4$.   (a)  Noiseless.   (b)
$\protect\sigma=0.05$.  The reported value of the correlation entropy is shown in \RVcolor.}
\end{figure}
\begin{figure*}[t!]
\begin{center}
    \subfloat[\label{fig:HAK-a}]{\includegraphics[width=0.35\columnwidth,keepaspectratio]{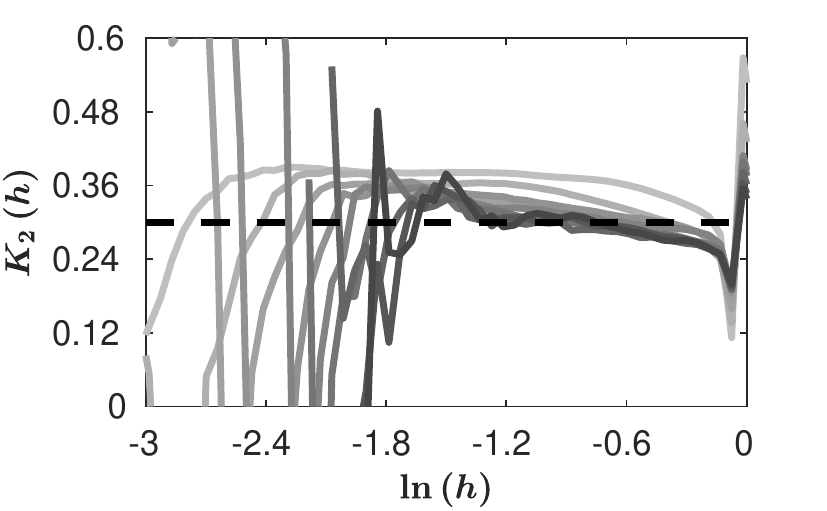}}
    \subfloat[\label{fig:HAK-b}]{\includegraphics[width=0.35\columnwidth,keepaspectratio]{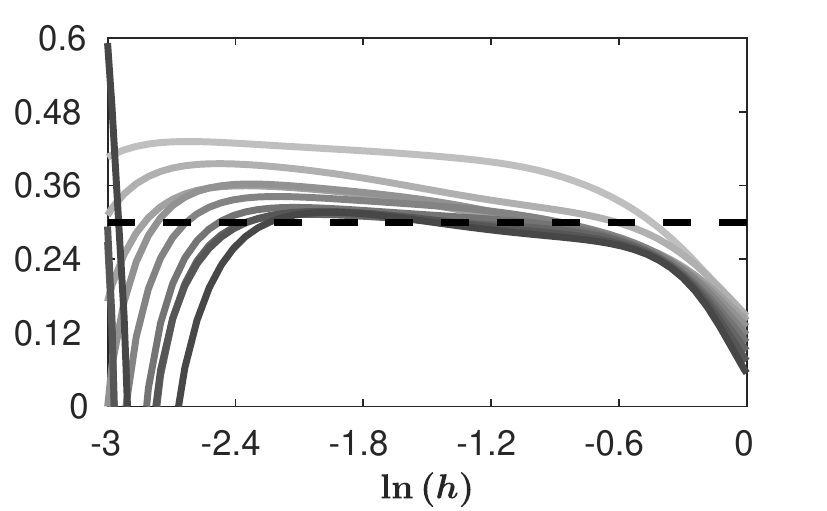}}
    \subfloat[\label{fig:HAK-c}]{\includegraphics[width=0.35\columnwidth,keepaspectratio]{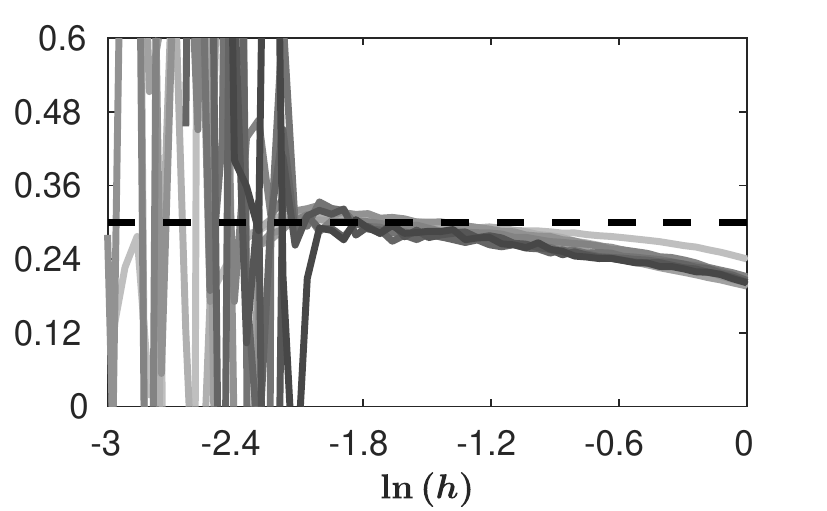}}\\
\end{center}
\caption{\label{fig:HenonAllK2} Henon map with noise  ($\sigma=0.05$).  Estimation of $K_2$ for $m=\W{4,6,\dots,22}$
through different coarse-grained estimators,  the curves for different $m$ values are color-coded in grayscale where
the lightest  gray corresponds to  $m=4$.  (a) Jayawardena~\glsname{etal} (b)~Nolte~\glsname{etal}  (c) $\CKu\E{h}$.
The reported value $K=0.3$ is shown in \RVcolor.}
\end{figure*}
\begin{figure*}[t!]
\begin{center}
    \subfloat[\label{fig:RAK-a}]{\includegraphics[width=0.35\columnwidth,keepaspectratio]{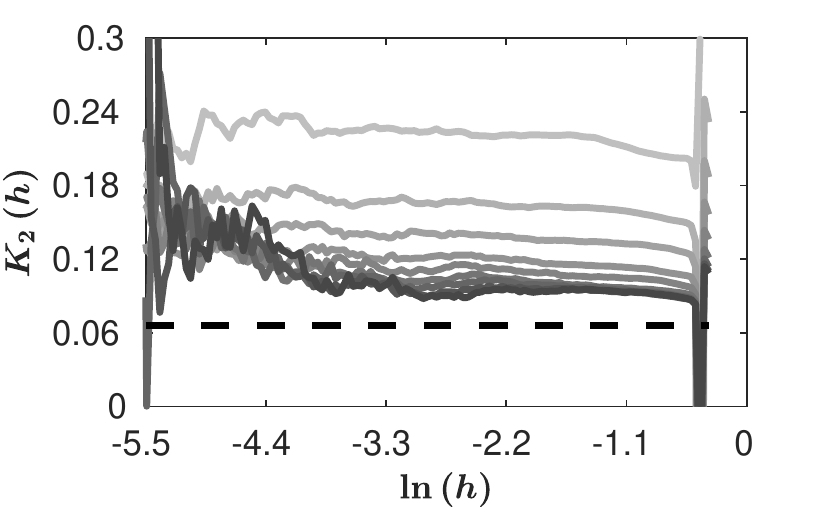}}
    \subfloat[\label{fig:RAK-b}]{\includegraphics[width=0.35\columnwidth,keepaspectratio]{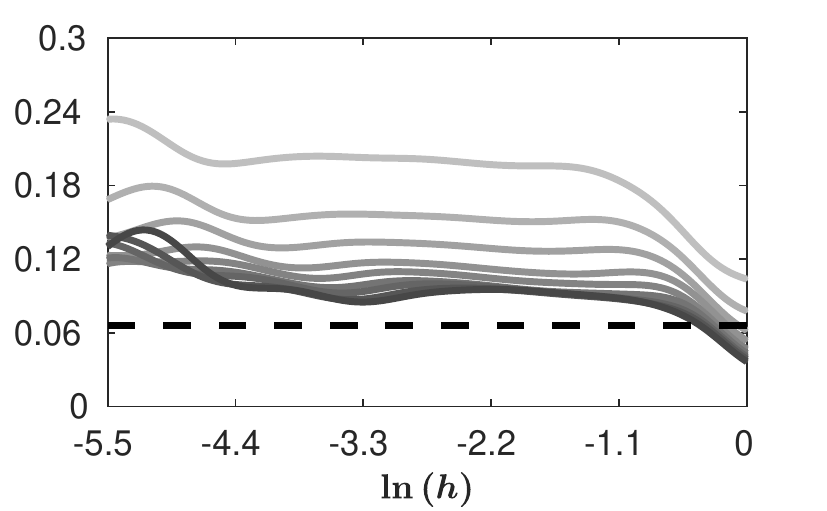}}
    \subfloat[\label{fig:RAK-c}]{\includegraphics[width=0.35\columnwidth,keepaspectratio]{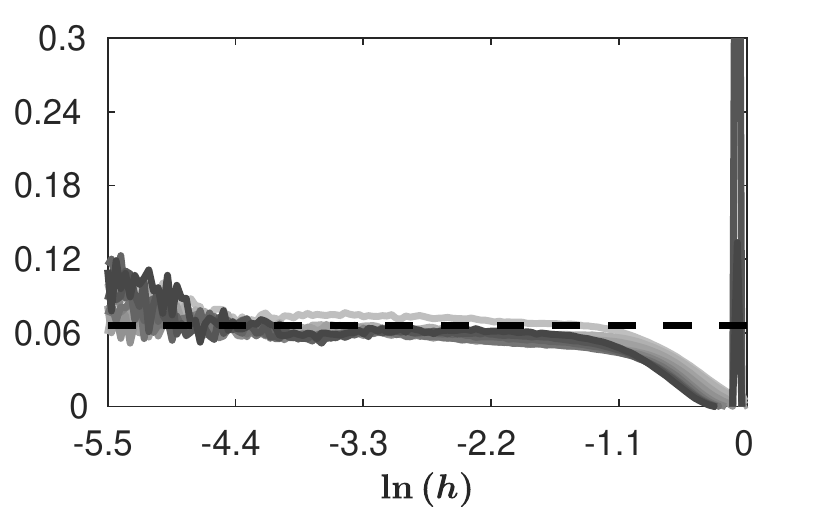}}\\
\end{center}
\caption{\label{fig:RosslerAllK2} Rössler  system.  Estimation of $K_2$  for $m=\W{4,6,\dots,22}$  through different
coarse-grained estimators,  the curves for different $m$ values are color-coded in grayscale where the lightest gray
corresponds to $m=4$.  (a) Jayawardena~\glsname{etal}  (b)~Nolte~\glsname{etal} (c) $\CKu\E{h}$.  The reported value
$K=0.066$ is shown in \RVcolor.}
\end{figure*}
\subsection{\label{sub:KE}Correlation entropy coarse-grained estimator}

Here proposed coarse-grained entropy estimator is defined as (see~Appendix~\ref{sec:A2}):  
\begin{align}
    \CKu\E{h}=&-\Frac{2\itau}\ln\E{\frac{\UCi{m+2}{\beta=m+2}\E{h}}{\UCi{m}{\beta=m}\E{h}}}\nonumber\\
              &-\Frac{2\itau}\ln\E{\frac{\CDu\E{h}}{m}+1}\nonumber\\
              &+\Frac{2\itau}\ln\Bigg[\Del\E{h}\E{\frac{m\m\logDm\E{h}}{m\m \CDu\E{h}}}\nonumber\\
              &+\E{1\m\Del\E{h}}\E{{\frac{\logDm\E{h}}{m}\p1}}\Bigg].\label{eq:CoarseK2}
\end{align}

This   coarse-grained   estimator   depends   on   the   correlation   integrals   $\UCi{m+2}{\beta=m+2}\E{h}$   and
$\UCi{m}{\beta=m}\E{h}$,  $\Del\E{h}$,  the logarithmic derivative  $\logDm\E{h}$,  and the coarse-grained estimator
$\CDu\E{h}$.  When $\sigma\ll  h$ the quantities $\CDu\E{h}\to\logDm\E{h}\to  D$,  and taking into  account that for
$\sigma\ll h$~\cite{Restrepo2016}:
\begin{equation*}
    \Frac{2\itau}\ln\Q{\frac{\UCi{m+2}{\beta=m+2}\E{h}}{\UCi{m}{\beta=m}\E{h}}}
    \approx\Frac{2\itau}\ln\E{\frac{D}{m}+1}-K_2, 
\end{equation*}
then it can be easily shown that: 
\begin{equation*}
    \CKu\E{h}\to K_2, \quad \text{for}\quad \sigma \ll h \quad \text{and}\quad m\gg D.
\end{equation*}

The estimations  of $\CKu\E{h}$  from both a  clean and a  noisy realization  ($\sigma=0.05$) of  the Henon  map for
$m=\W{4,6,8}$  are presented  in Fig.~\ref{fig:Ku}.  It  can be  observed that  in both  cases (Fig.~\ref{fig:Ku-a},
Fig.~\ref{fig:Ku-b}) $\CKu\E{h}$ is  near to the reported value  of $K_2=0.3$ for all $m$  values.  This is the main
advantage of the \gls{uci} in comparison with other correlation integrals~\cite{Restrepo2016}.

The last  statement can be confirmed  by comparing $\CKu\E{h}$ with  two similar coarse-grained  estimators based on
different correlation  integrals.  The first one was  proposed by Jayawardena~\glsname{etal}~\cite{Jayawardena2010},
and it is based  on the Grassberger and Procaccia correlation integral  ($C_{m}\E{h}$).  The second one was proposed
by Nolte~\glsname{etal},  and it  is calculated from the \gls{gci} ($T_m\E{h}$).  These  approaches,  as well as the
here proposed, $\CKu\E{h}$ only requires the estimation of two correlation integrals of the same kind.

In  Fig.~\ref{fig:HenonAllK2}, the  estimations of  \gls{k2} for  the Henon  map through  the estimators  proposed by
Jayawardena~\glsname{etal}, Nolte~\glsname{etal} and $\CKu\E{h}$ can be compared.  They were calculated from a time
series  of  length $N=5000$  with  added white  Gaussian  noise  at  a  level  $\sigma=0.05$.  The  correlation sums
$C_{m}\E{h}$,    $T_m\E{h}$,    $\UCi{m}{\beta=m}\E{h}$   and   $\UCi{m}{\beta=m-2}\E{h}$    were   calculated   for
$m=\{4,6,\dots,22\}$  with  $\tau=1$.   Figure~\ref{fig:HAK-a}  shows  the  results  for  the  estimator  proposed  by
Jayawardena~\glsname{etal} It can be  observed that,  as $m$ is increased,  this estimator  slowly approaches to the
reported value of \gls{k2}.  However,  it must be pointed out that, regardless that the dimension of this map is low
($D=1.22$),  high  values  of  $m$ are  needed  in  order to  converge.  On  the  other  hand,  it  can  be  seen in
Fig.~\ref{fig:HAK-b}  that  the  results of  the  coarse-grained  estimator  proposed  by  Nolte~\glsname{etal} also
converge slowly to the reported value for \gls{k2}.  In contrast,  in Fig.~\ref{fig:HAK-c} it can be observed
that the  estimator $\CKu\E{h}$ approaches faster  to the value of  \gls{k2} than the other  two studied estimators.
Note that  the curves  for different $m$  values of $\CKu\E{h}$  are very  close to  each other,  meaning  that this
estimator is more robust to changes in the parameter $m$.  Please note that the curves corresponding to $m=\W{4,  6,
8}$ of Fig.~\ref{fig:HAK-c} are also shown in Fig.~\ref{fig:Ku-b}.

The same methodology was applied to a clean time series obtained from the Rössler system:

\begin{equation*}
\begin{cases}\dot{x}&=-y-z\\\dot{y}&=x+ay\\\dot{z}&=b+z\E{x-c}\end{cases},
\end{equation*}
where $a=0.15$,  $b=0.2$ and  $c=10$.  The system was solved using  the Runge-Kutta method with time  step $0.5$ and
initial condition $x\E{0}=0.1$,  $y\E{0}=0.1$,  $z\E{0}=0.1$.  The first $20000$ data points were discarded, and the
next $N=10000$ data  points were taken to  estimate the attractor's invariants.  The  correlation sums $C_{m}\E{h}$,
$T_m\E{h}$,  $\UCi{m}{\beta=m}\E{h}$ and  $\UCi{m}{\beta=m-2}\E{h}$  were  calculated  for $m=\W{4,6,\dots,22}$ with
$\tau=3$.

The   results   for   the   estimators   proposed   by   Jayawardena,   Nolte,   and   $\CKu\E{h}$   are   shown  in
Figs.~\ref{fig:RAK-a},~\ref{fig:RAK-b} and~\ref{fig:RAK-c},  respectively.  In  all three cases,  a  typical scaling
behavior can be observed.  Once again,  the estimators proposed by Jayawardena and Nolte need high values of $m$, in
contrast with  $\CKu\E{h}$.  Interestingly,  all estimators converge to  values of correlation  entropy that differs
from the  one reported  ($K_2=0.066$)~\cite{Diks1999}.  Both the  Jayawardena's and  Nolte's estimators  converge to
$\hat{K}_2\approx0.085$,  while  $\CKu\E{h}$  converges  to  $\hat{K}_2\approx0.052$.  From  these  results,  we can
conclude that  the estimator $\CKu\E{h}$ converges  by using low values  of $m$,  in comparison  with the previously
proposed estimators~\cite{Nolte2001,Jayawardena2010}.

\section{\label{sec:AIS}Automatic invariants estimation}

In practical applications,  it is important  to verify the existence of a scaling behavior  for a range of $h$.  The
determination of a stable scaling region,  which is required  to estimate the invariants,  is usually done by visual
inspection.  However, that is not feasible when large sets of data are studied Additionally, the subjective judgment
should be  avoided.  The necessity  of an  algorithm for the  automatic estimation  of invariants  arises,  once the
scaling regime has been confirmed.  In this section, we propose an algorithm (see algorithm~\ref{alg:AEAI}) based on
the coarse-grained estimators $\CSu\E{h}$, $\CDu\E{h}$ and $\CKu\E{h}$, that is able to automatically estimate these
invariants.  Before  describing this  algorithm,  we need  to detail  some  aspects  about  the  calculation  of the
U-correlation integral.

\subsection{\label{sub:NAA}Noise-assisted correlation algorithm}

\begin{figure}[t!]
\begin{center}
    \includegraphics[width=0.5\columnwidth,keepaspectratio]{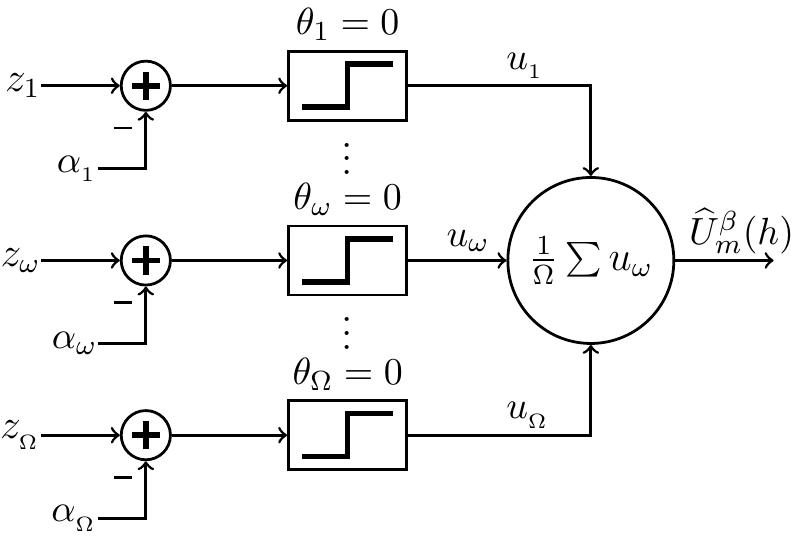}
\end{center}
\caption{\label{fig:NAA}Noise-assisted       correlation      algorithm.       The       $\omega$-th      comparator
($\omega=1,2,\dots,\Omega$) with  threshold ${\theta_\omega=0}$ receives as  input the squared  distance between two
$m$-dimensional delay  vectors  minus  an  i.i.d.  realization $\alpha_\omega~\sim\E{h^2/2}\inlineChiS$.  All binary
outputs $\mu_\omega$ are averaged to calculate the U-correlation sum $\widehat{U}_{m}^{\beta}\E{h}$.}
\end{figure}

The  \gls{uci}  is  approximated  with  U-correlation  sum  $\widehat{U}_{m}^{\beta}\!\E{h}$.   The  calculation  of
$\widehat{U}_{m}^{\beta}\!\E{h}$ requires a computationally expensive numerical integration due to the kernel of the
\gls{uci}.   However,   we  can  use  the  noise-assisted  correlation  algorithm  instead~\cite{Restrepo2016}  (see
Fig.~\ref{fig:NAA}).  This algorithm begins by calculating the squared  distances between each pair of delay vectors
$z_{\omega}=\norm{\vv{x_i}-\vv{x_j}}^2$,       where       $\omega=\{\left(i,j\right)/\,\,\,i=1,2,\dots       L,\,\,
j=1,2,\dots,L,\,\,\,i\neq j\}$ and $\Omega=L\E{L-1}$.  Then, for each squared distance $z_{\omega}$ an i.i.d.  noise
realization $\alpha_{\omega}\sim\text{Gamma}\E{\beta/2,h^2}$ is subtracted.  Here, $\beta/2$ and $h^2$ are the shape
and scale parameters,  respectively.  The result is then compared  with a threshold $\theta_{\omega}=0$ to produce a
binary  output $u_{\omega}$.  Then  $\widehat{U}_{m}^{\beta}\!\E{h}$ is  the average  of all  binary outputs.  These
steps must be repeated for each value of $h$  and $m$.  Algorithm~\ref{alg:NCA} describes all the steps that must be
followed in order  to estimate the noise-assisted U-correlation sum.  The  noise realizations $\alpha_{\omega}$ were
generated     using     $\inlineChiS$     realizations     given     that,      if     $X\sim\inlineChiS$,      then
$\E{h^2/2}X\sim\text{Gamma}\E{\beta/2,h^2}$.

As    an    example,    the    correlation    integral    $\UCi{m}{\beta=m}\E{h}$    can    be    approximated    by
$\widehat{U}_{m}^{\beta=m}\!\E{h}$  which  in  turn  is  calculated  from  the  squared  distances  $z_{\omega}$  of
$m$-dimensional delay vectors,  and  noise realizations $\alpha_\omega$ that follow a  Gamma distribution with shape
parameter equal  to $m/2$.  Similarly,  the correlation  sum $\widehat{U}_{m+2}^{\beta=m-2}\!\E{h}$  can be
obtained  from the  squared distances  $z_{\omega}$ of  $(m+2)$-dimensional  delay  vectors  and  noise realizations
$\alpha_\omega$ following a Gamma distribution with shape parameter equal to $m/2-1$.
\begin{algorithm}[t!]
  \caption{\label{alg:NCA}Noise-assisted U-correlation algorithm.}
   \begin{algorithmic}[1]
    \State~Form the  $m$-dimensional delay  vectors from  the temporal  series and  calculate all  pairwise squared
    distances $z_\omega$ with $1\leq\omega\leq \Omega=L\E{L-1}$.%
    \State~Fix the  value of the parameter $h$  and obtain $\Omega$ realization of  noise $\alpha_{\omega}$ from the
    distribution $\E{h^2/2}\inlineChiS$.%
    \State~Calculate the binary variable:%
    \begin{equation*}
        u_{\omega}\E{h}=\begin{cases} 0 &\text{if}\quad z_\omega-\alpha_\omega\geq0 \\ 1& \text{otherwise}\end{cases}.
    \end{equation*}
    \State~Calculate the  U-correlation sum as:%
    \begin{equation*}
        \widehat{U}_{m}^{\beta}\E{h}=\Frac{\Omega}\sum_{\omega=1}^{\Omega}{u_{\omega}}.
    \end{equation*}%
    \State~Repeat steps $1$-$4$ for all values of $h$ and $m$.%
   \end{algorithmic}
\end{algorithm}
\subsection{\label{sub:AUTO}Algorithm for automatic estimation of invariants}

\begin{figure*}[t]
\begin{center}
    \subfloat[\label{fig:HS-a}]{\includegraphics[width=0.33\columnwidth,keepaspectratio]{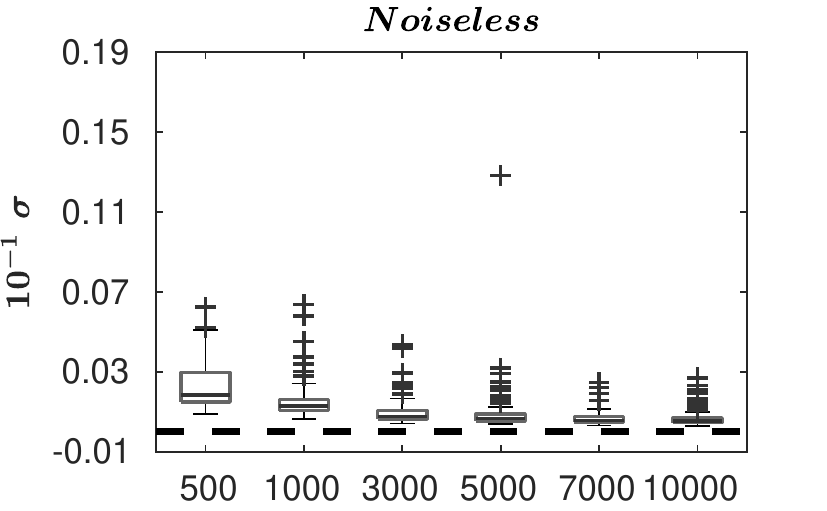}}
    \subfloat[\label{fig:HS-b}]{\includegraphics[width=0.33\columnwidth,keepaspectratio]{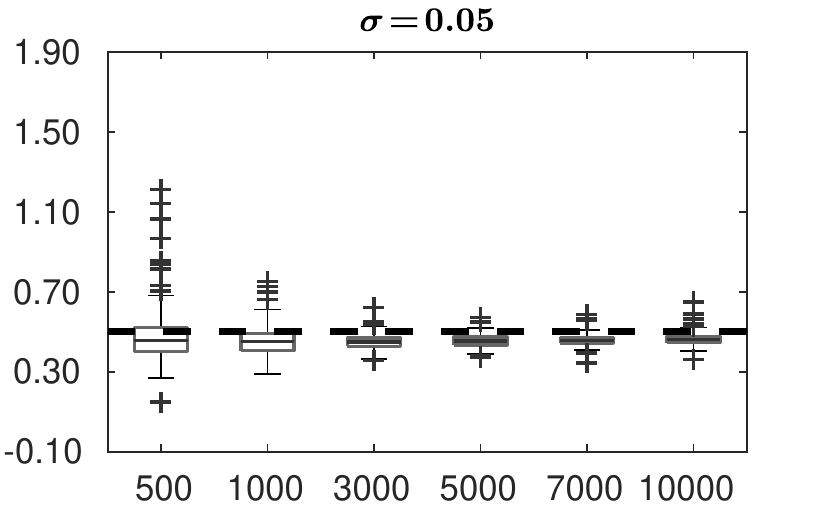}}
    \subfloat[\label{fig:HS-c}]{\includegraphics[width=0.33\columnwidth,keepaspectratio]{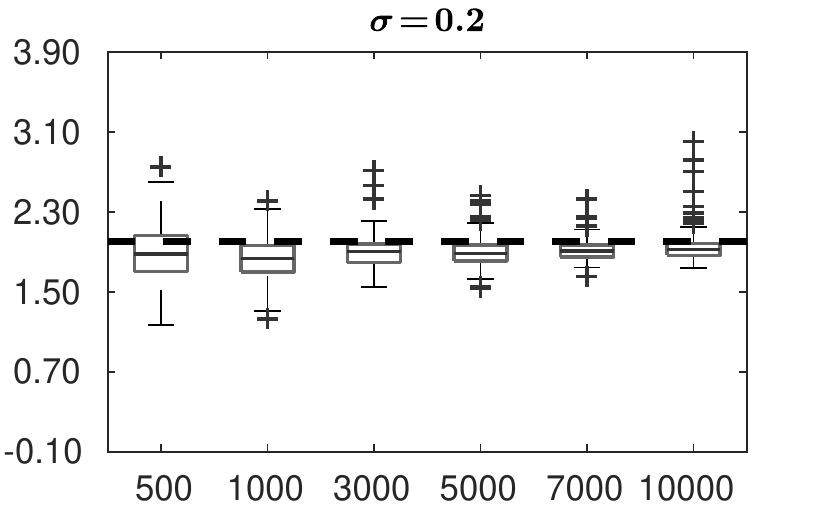}}\\
    \subfloat[\label{fig:HD-a}]{\includegraphics[width=0.33\columnwidth,keepaspectratio]{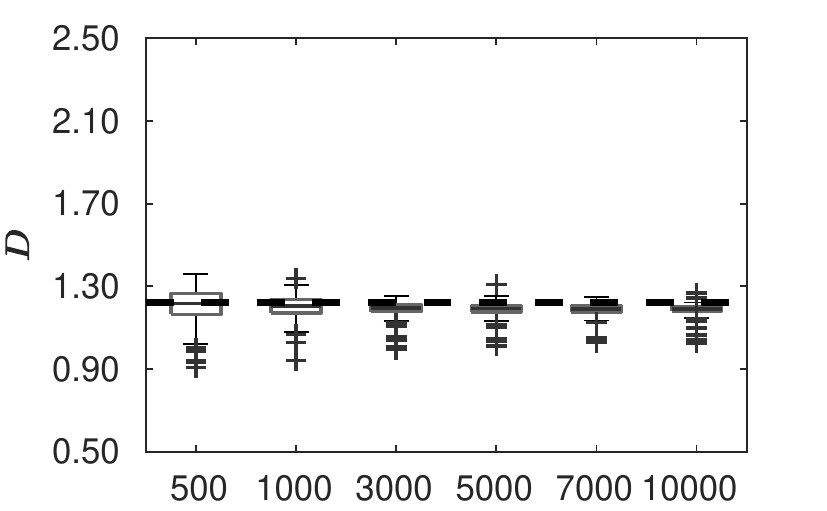}}
    \subfloat[\label{fig:HD-b}]{\includegraphics[width=0.33\columnwidth,keepaspectratio]{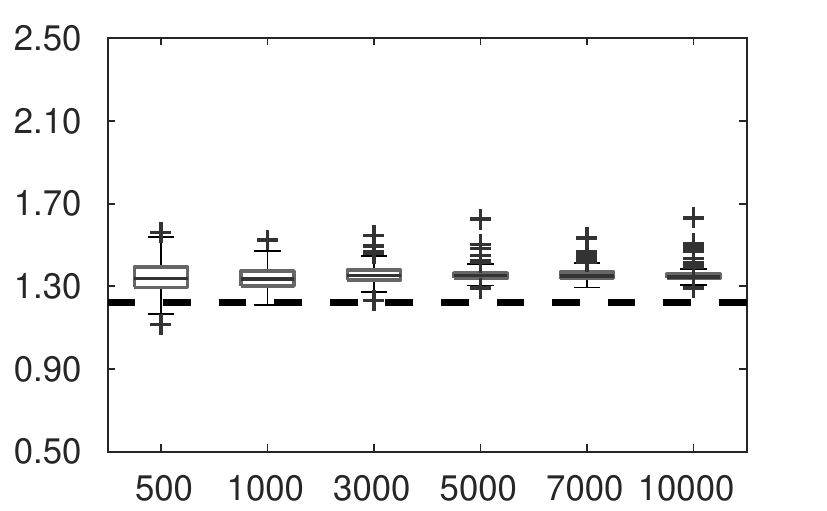}}
    \subfloat[\label{fig:HD-c}]{\includegraphics[width=0.33\columnwidth,keepaspectratio]{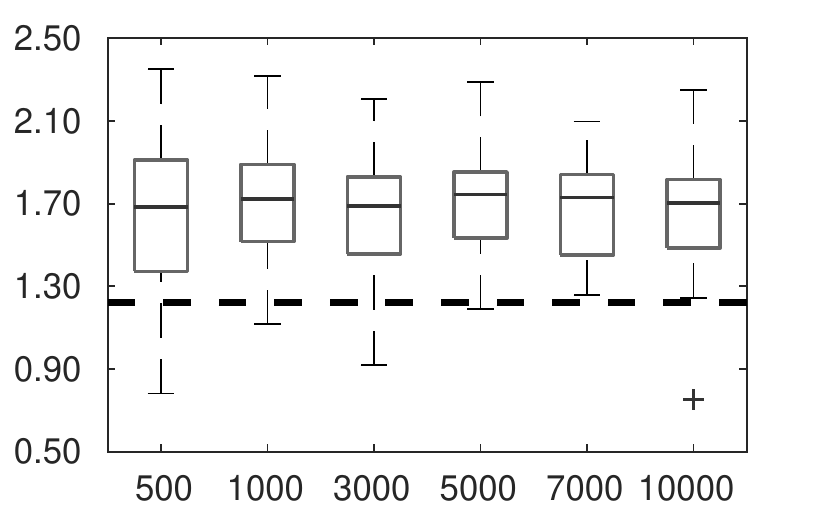}}\\
    \subfloat[\label{fig:HK-a}]{\includegraphics[width=0.33\columnwidth,keepaspectratio]{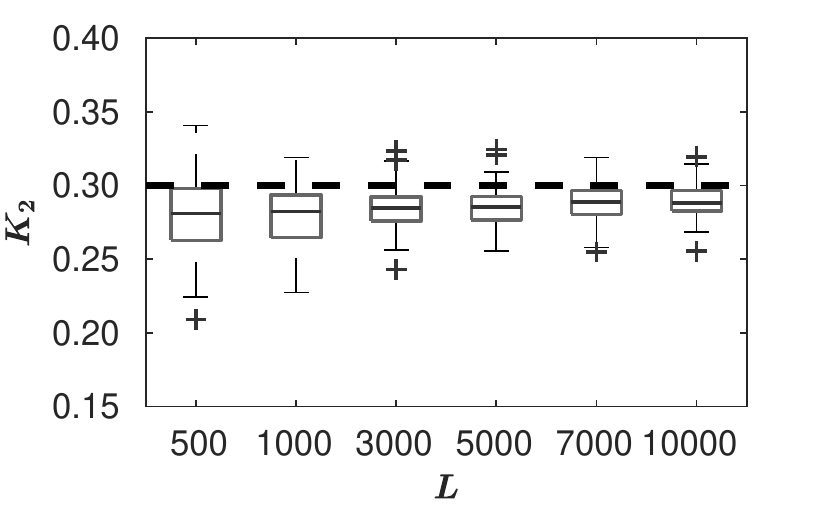}}
    \subfloat[\label{fig:HK-b}]{\includegraphics[width=0.33\columnwidth,keepaspectratio]{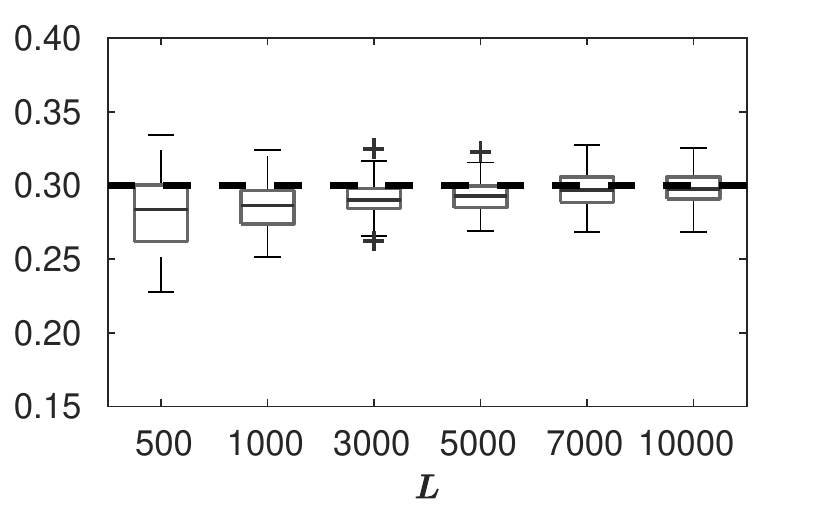}}
    \subfloat[\label{fig:HK-c}]{\includegraphics[width=0.33\columnwidth,keepaspectratio]{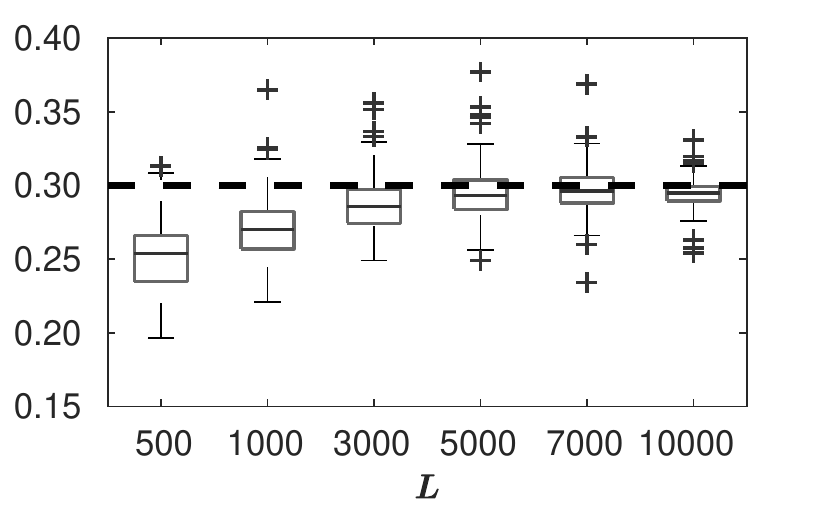}}
\end{center}
  \caption{\label{fig:StatHenon}Henon map.  Box plot  of the estimation of $\sigma$,  $D$,  and  $K_2$ for different
  numbers  of  available  delay  vectors  ${L=\W{500,1000,3000,5000,10000}}$.  First  row:  estimation  of $\sigma$.
  (a)~noiseless,   (b)~$\sigma=0.05$,   and  (c)~$\sigma=0.2$.   Second  row:  estimation  of  $D$.   (d)~noiseless,
  (e)~$\sigma=0.05$,  and (f)~$\sigma=0.2$.  Third row: estimation of $K_2$.  (g)~noiseless,  (h)~$\sigma=0.05$, and
  (i)~$\sigma=0.2$.  The reported values are shown in \RVcolor.}
\end{figure*}

The  automatic  algorithm   to   estimate   invariants   begins   by   approximating   the  U-correlation  integrals
$\UCi{m}{\beta=m}\E{h}$ and  $\UCi{m}{\beta=m-2}\E{h}$,  for different $m$  values ($m>2$) using  the noise-assisted
correlation algorithm  (Alg.~\ref{alg:NCA}).  This step does  not requires a narrow  range of $h$  as initial guess.
For example, for a normalized time series the range $10^{-8}\leq h\leq10$ is a good choice.  On the other hand, note
that the  expression $\UCi{m}{\beta=m-2}\E{h}$ requires values  of $m>2$ since Gamma  distribution's shape parameter
must be greater than zero.

Next,  the logarithmic  derivatives  $\logDm\E{h}$,  $\logDmb{\beta=m}{m+2}\E{h}$  are  computed  and $\Del\E{h}$ is
calculated using Eq.~(\ref{eq:delta}).  To calculate the logarithmic derivatives, we make use of a wavelet transform
approach  to  approximate the  derivative~\cite{Luo2006}.  This  allows us  to  achieve  better  estimations  of the
invariants because  it implements  a low-pass  filter reducing  the high-frequency  oscillations that  are naturally
present  in  estimators  derived  from   the  noise-assisted  algorithm~\cite{Restrepo2016}.   Please  observe  that
correlation  integral  $\UCi{m+2}{\beta=m}\E{h}$  is   equal  to  $\UCi{\hat{m}}{\beta=\hat{m}-2}\E{h}$  correlation
integral evaluated at $\hat{m}=m+2$.  In order to obtain \gls{nl},  the coarse-grained estimator $\CSu\E{h}$ must be
calculated [Eq.~(\ref{eq:CoarseS})].

To obtain a suitable range of scale values where the  invariants can be estimated,  one must look for a range of $h$
where the  coarse-grained estimator is  nearly constant and  its variation across  the different  $m$ values  is the
smallest.  In this sense, for the noise level we define the functions:
\begin{align}
    A_{\sigma}\E{h}=& \Frac{M}\sum_{i=1}^{M}\dif{\,\sigma^{U}_{m_{i}}\E{h}}{\ln h},\nonumber \\
    V_{\sigma}\E{h}=&\Frac{M-1}\sum_{i=1}^{M}{\E{\sigma^{U}_{m_{i}}\E{h}-\hat{\sigma}^{U}\E{h}}^2},\nonumber\\
    \intertext{and}
    F_{\sigma}\E{h}=&A_{\sigma}\E{h} V_{\sigma}\E{h},\nonumber
\end{align}
where $m\in\W{m_1,m_2,\dots,m_i,\dots,m_M}$,  $\hat{\sigma}^{U}\E{h}$  is  the  average  of  $\CSu\E{h}$ across $m$.
$A_{\sigma}\E{h}$ is the  average over $m$ of the  derivative of $\CSu\E{h}$ with respect  to $\ln h$,  $V_{\sigma}$
gives the variation of $\CSu\E{h}$ across $m$ and  $F_{\sigma}\E{h}$ is the product of the aforementioned functions.
We propose  to estimate \gls{nl}  within a range  of $h$ centered  at the  $h$ value  at which  $F_{\sigma}\E{h}$ is
minimum.  In this way,  \gls{nl} is estimated  in a range of $h$ centered in a  plateaued region (scaling region) of
$\CSu\E{h}$, and its value is consistent through the parameter $m$.

The correlation dimension is determined using the coarse-grained estimator $\CDu\E{h}$ [Eq.~(\ref{eq:CoarseD})].  To
find a range of $h$ values to estimate \gls{cd}, we define the functions $A_{D}\E{h}$,  $V_{D}\E{h}$ and $F_{D}\E{h}$
similarly to the ones above but using the  estimator $\CDu\E{h}$ instead of $\CSu\E{h}$.  The estimation of \gls{cd}
must be done within a range of $h$ centered at the $h$ value at which $F_{D}\E{h}$ is minimum.

Finally, the correlation entropy can be estimated using the coarse-grained estimator $\CKu\E{h}$.  As it can be seen
in Eq.~(\ref{eq:CoarseK2}),  this  estimator depends  on $\Del\E{h}$,  $\logDm\E{h}$,  the  coarse-grained estimator
$\CDu\E{h}$  and  the  ratio  between  $\UCi{m}{\beta=m}\E{h}$  and  $\UCi{m+2}{\beta=m+2}\E{h}$.  \gls{k2}  must be
determined within a range of $h$ centered at the $h$ value at which $F_{K_2}\E{h}$ is minimum.

\begin{algorithm}[t!]
  \caption{\label{alg:AEAI}Automatic estimation of attractors' invariants.}
   \begin{algorithmic}[1]
        \State~Calculate $\UCi{m}{\beta=m}\E{h}$ and $\UCi{m}{\beta=m-2}\E{h}$ using Alg.~\ref{alg:NCA} for $m>2$.
        \State~Calculate $\Del\E{h}$ using Eq.~(\ref{eq:delta}) and the \gls{uci} obtained in step $1$.
        \State~Compute $\CSu\E{h}$ with Eq.~(\ref{eq:CoarseS})  and obtain the function $F_{\sigma}\E{h}$.  Estimate
        $\sigma$ within a range of $h$ centered at the value $h$ where $F_{\sigma}\E{h}$ is minimum.
        \State~Use $\UCi{m}{\beta=m}\E{h}$,  $\UCi{m}{\beta=m-2}\E{h}$  and  $\Del\E{h}$  to  calculate $\CDu\E{h}$
        (Eq.~(\ref{eq:CoarseD})) and obtain the function $F_{D}\E{h}$.  Estimate  $D$ within a range of $h$ centered
        at the value $h$ where $F_{D}\E{h}$\footnotemark~is minimum.
        \State~Obtain   $\CKu\E{h}$  (Eq.~(\ref{eq:CoarseK2}))  using   $\UCi{m}{\beta=m}\E{h}$,   $\Del\E{h}$  and
        $\CDu\E{h}$.  Calculate the function  $F_{K_2}\E{h}$.  Estimate $K_2$ within a range of  $h$ centered at the
        value $h$ where $F_{D}\E{h}$ is minimum.
   \end{algorithmic}
\end{algorithm}
\footnotetext{The functions $F_{D}\E{h}$ and $F_{K_2}\E{h}$ can be computed similarly to $F_{\sigma}\E{h}$
but using the coarse-grained estimators $\CDu\E{h}$ and $\CKu\E{h}$, respectively.}
\subsection{\label{sub:Sim}Simulations}

The aforementioned  methodology was applied to  a set of $128$  realization of  the Henon  map with  different noise
levels  ${\sigma=\W{0,0.05,0.2}}$ and  data lengths  which are  proportional to  the number  $L$ of  available delay
vectors  to  estimate   the  invariants  \gls{nl},   \gls{cd}  and  \gls{k2}.   To   calculate  the  \gls{uci}s  for
${m=\W{4,5,\dots,8}}$ all realizations were normalized (unitary  standard deviation),  and the nearest $15$ temporal
neighbors of each delay vector were discarded~\cite{Kantz2004}.
\begin{figure*}[t]
\begin{center}
    \subfloat[\label{fig:MS-a}]{\includegraphics[width=0.33\columnwidth,keepaspectratio]{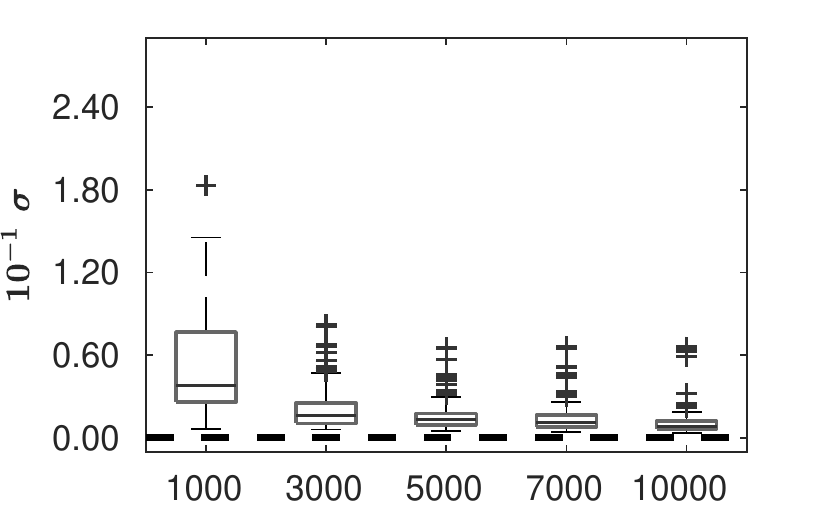}}
    \subfloat[\label{fig:MS-b}]{\includegraphics[width=0.33\columnwidth,keepaspectratio]{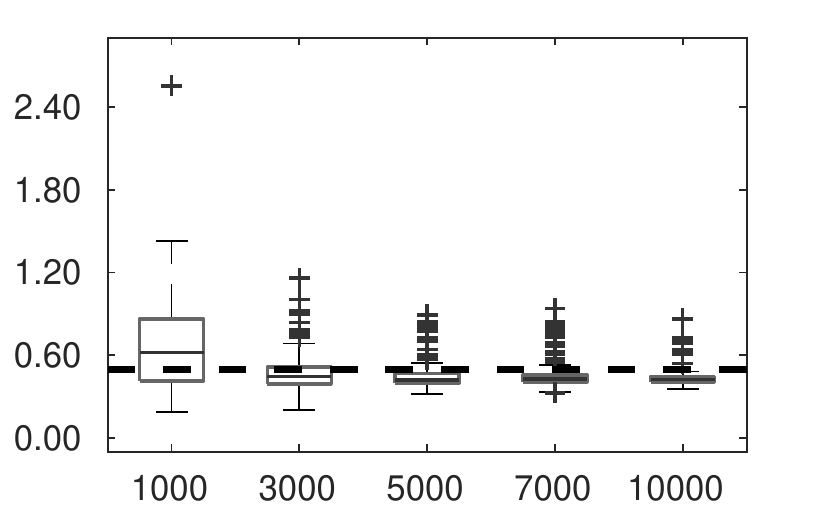}}
    \subfloat[\label{fig:MS-c}]{\includegraphics[width=0.33\columnwidth,keepaspectratio]{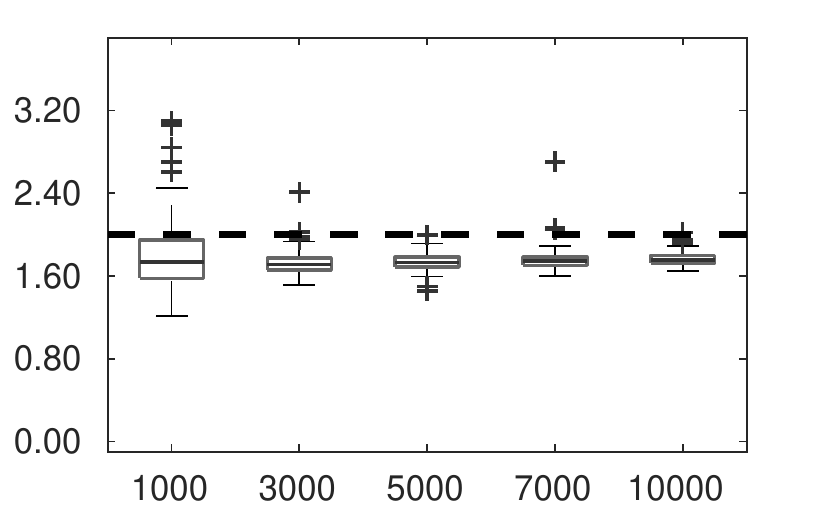}}\\
    \subfloat[\label{fig:MD-a}]{\includegraphics[width=0.33\columnwidth,keepaspectratio]{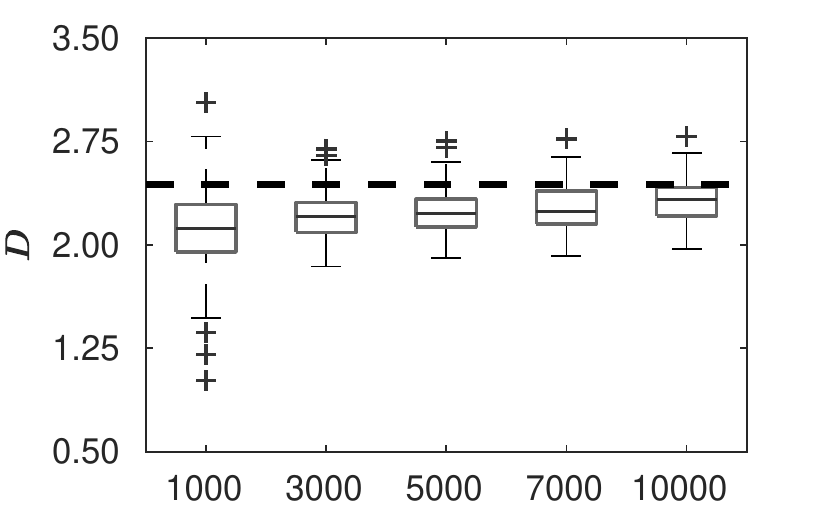}}
    \subfloat[\label{fig:MD-b}]{\includegraphics[width=0.33\columnwidth,keepaspectratio]{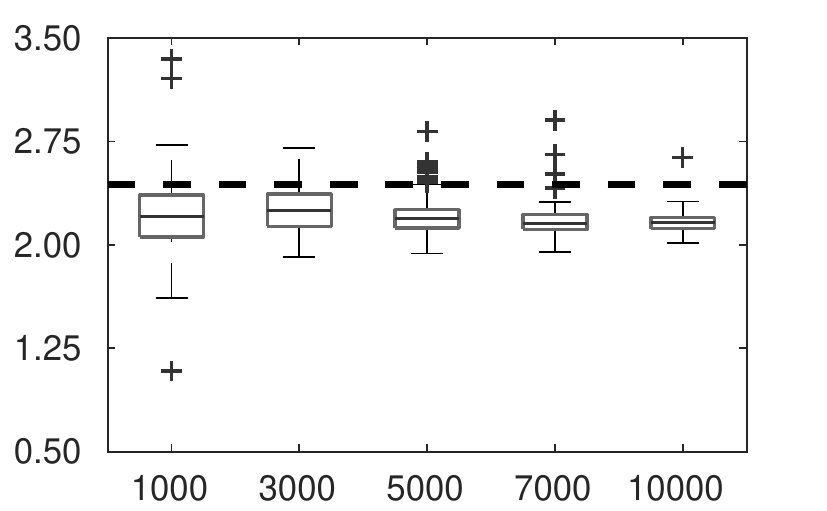}}
    \subfloat[\label{fig:MD-c}]{\includegraphics[width=0.33\columnwidth,keepaspectratio]{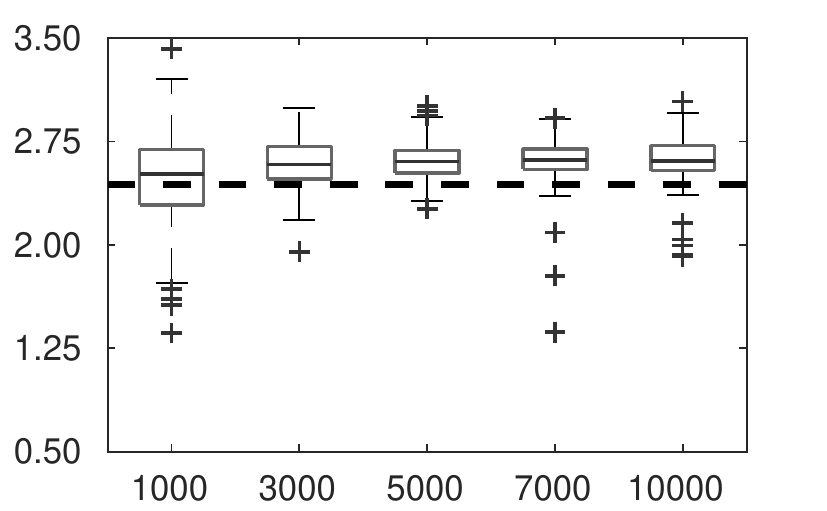}}\\
    \subfloat[\label{fig:MK-a}]{\includegraphics[width=0.33\columnwidth,keepaspectratio]{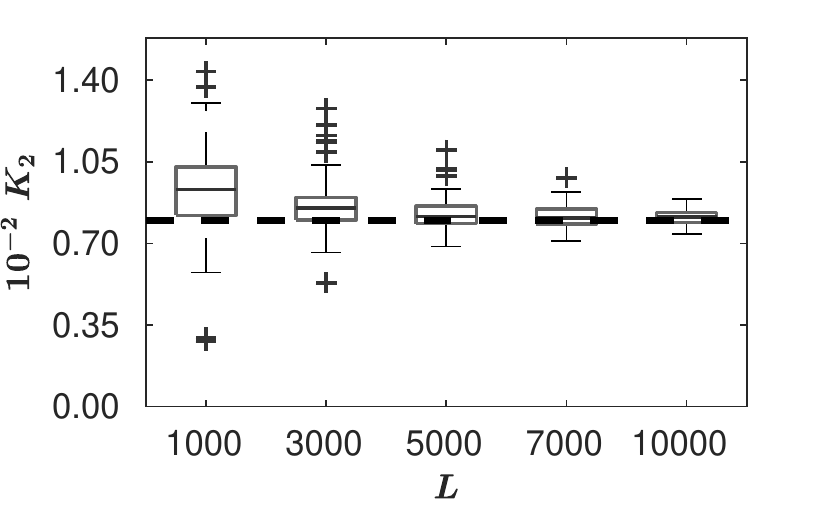}}
    \subfloat[\label{fig:MK-b}]{\includegraphics[width=0.33\columnwidth,keepaspectratio]{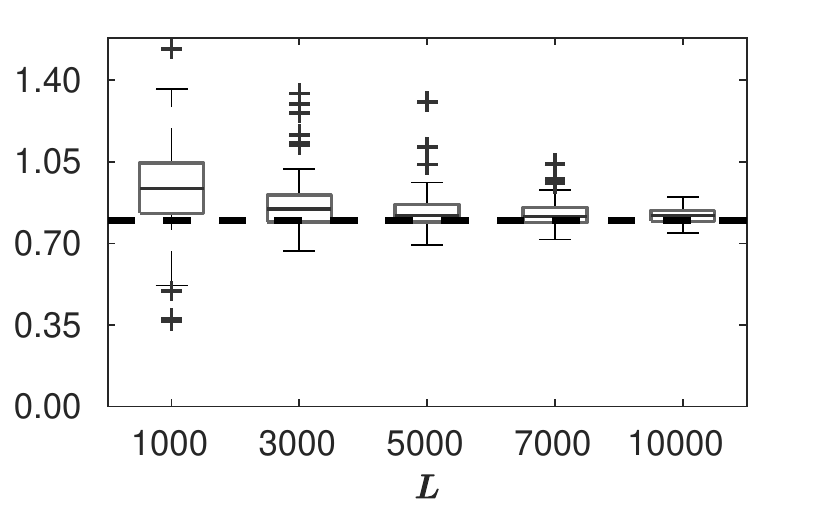}}
    \subfloat[\label{fig:MK-c}]{\includegraphics[width=0.33\columnwidth,keepaspectratio]{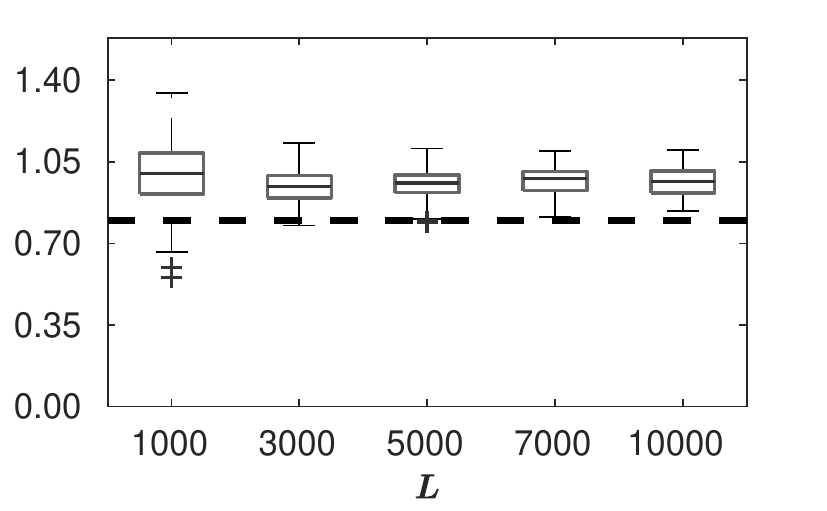}}
\end{center}
 \caption{\label{fig:StatMack} Mackey-Glass box  plot of the estimation of  $\sigma$,  $D$,  and $K_2$ for different
 numbers  of  available  delay  vectors,   ${L=\W{1000,3000,5000,10000}}$.   First  row:   estimation  of  $\sigma$.
 (a)~noiseless,   (b)~$\sigma=0.05$,   and  (c)~$\sigma=0.2$.   Second  row:   estimation  of  $D$.   (d)~noiseless,
 (e)~$\sigma=0.05$,  and (f)~$\sigma=0.2$.  Third row:  estimation of $K_2$.  (g)~noiseless,  (h)~$\sigma=0.05$, and
 (i)~$\sigma=0.2$.  The reported values are shown in \RVcolor.}
\end{figure*}
Figure~\ref{fig:StatHenon}   shows  the   boxplots  of   the  invariants   \gls{nl},   \gls{cd}  and   \gls{k2}  for
$L=\W{500,1000,3000,5000,10000}$.  Unfortunately,  there  is no  other automatic  algorithm  for  the  estimation of
invariants.  Due to this reason,  we only can compare the results of our estimations with previously reported values
calculated by  selecting the  scaling range  by visual  inspection.  For the  noise level  estimation (first  row of
Fig.~\ref{fig:StatHenon}),  it can be observed that,  in absence of noise,  our algorithm overestimates the value of
\gls{nl} (\RVcolor).  On the contrary,  for higher noise levels ($\sigma=\W{0.05,0.2}$) the proposed method slightly
underestimates \gls{nl}.  It can be noticed that,  regardless the noise level,  the median of each box is very close
to the true value of  \gls{nl} and the variance of the estimation decreases as  the number of data points increases.
The estimation of the correlation dimension is shown  in the second row of Fig.~\ref{fig:StatHenon}.  It can be seen
in Fig.~\ref{fig:HD-a} that for noiseless time series the estimation of \gls{cd} is very close to the reported value
($D=1.22$) and its variance decreases as the length  of the time series is increased.  For higher noise levels,  the
correlation dimension is slightly biased, but the estimations are still close to the previously reported value.

The results for  the correlation entropy are  shown in the third row  of Fig.~\ref{fig:StatHenon}.  In general,  the
here proposed method  yields correlation entropy values converging  to values lesser than the  reported for this map
($K_2=0.3$).

The proposed approach was also tested on the Mackey-Glass time series, which is generated by the following nonlinear
time-delay differential equation: \begin{equation*} \dot{x}\E{t} = \frac{ax\E{t-\lambda}}{1+\Q{x\E{t-\lambda}}^{10}}
- bx\E{t},\label{eq:Mackeyeq} \end{equation*} where $a=0.2$,  $b=0.1$, and $\lambda=23$.  We estimate the invariants
$\sigma$,  $D$,  and $K_2$ from $128$ realizations with different  initial conditions and normalized to have unitary
standard deviation.  The embedding dimensions were $m=\W{4,6,\dots,10}$ and  the embedding lag $\tau$ was set to the
first local minimum of the mutual information function ($\tau=20$)~\cite{Kaffashi2008a}.  Finally,  the nearest $15$
temporal neighbors of each delay vector were discarded.

The results of these simulations are presented in Fig.~\ref{fig:StatMack} and the estimations of the noise level are
presented in Figs.~\ref{fig:MS-a},  \ref{fig:MS-b} and \ref{fig:MS-c}.  In  absence of noise the algorithm estimates
small positive values of \gls{nl}.  This behavior is explained because the function $\Del\E{h}$ tends to zero as the
length of the time series increases,  producing more  accurate estimations with less variance.  In presence of noise
($\sigma=0.05$  and   $\sigma=0.2$)  the  here  proposed   algorithm  underestimates   the  noise   level,   but  in
Fig.\ref{fig:StatHenon} it can be seen how the median values tend to the real value of \gls{nl} as $L$ increases.

The estimations of the correlation dimension are shown in the second row of Fig.~\ref{fig:StatMack}.  For clean time
series,    our   algorithm    provides   estimations    of   \gls{cd}    very   close    to   the    reported   value
$D=2.44$~\cite{Grassberger1983a}.  As the data length is increased, the estimations approach the reported value.  In
case of  noisy time series,  the  estimations of the  correlation dimension are  greater than those  obtained in the
noiseless case.  Nevertheless, the estimation of \gls{cd} is still very good.

In  Figs.~\ref{fig:MK-a},  \ref{fig:MK-b}  and  \ref{fig:MK-c}  the  estimations  for  the  correlation  entropy are
presented.     Our    algorithm     achieves    satisfactory    estimations    of     \gls{k2}    (reported    value
$K_2=0.008$)~\cite{Grassberger1984},  even in presence  of noise.  It can be observed that regardless  the value of $L$
the estimation of $K_2$ is consistent, but its variance decreases as the data length increases.

\section{\label{sec:Dis}Discussion}

In this section,  we review  the main contributions of this article and give  some practical recommendations for its
implementation.  In this work, we introduce a new methodology to estimate the correlation dimension, the correlation
entropy and the  noise level of time series.  This  approach is based on the  recent proposed U-correlation integral
(Eq.~(\ref{eq:uci1})), introduced in~\cite{Restrepo2016}, which uses a kernel function that depends on the embedding
dimension.  From  the  scaling  function  of  the  \gls{uci}  (Eq.~(\ref{eq:UCI})),  we  have  developed  a  set  of
coarse-grained  estimators  for  the   correlation  dimension  (Eq.~(\ref{eq:CoarseD})),   the  correlation  entropy
(Eq.~(\ref{eq:CoarseK2}))  and the  noise level  (Eq.~(\ref{eq:CoarseS})).  The  advantage  of  these coarse-grained
estimators is that they only depend on the  estimation of two U-correlation integrals.  In other words,  they do not
need  of the  tuning of  any external  parameter and/or  the  estimation  of  the  noise  level,  as  it  does other
approaches~\cite{Diks1999,Jayawardena2008,Restrepo2016}.

From   the   results   presented   in  the   previous   section,   we   can   conclude   that   the   here  proposed
algorithm~(Alg.~\ref{alg:AEAI}) is able to  automatically select the scaling region and  to estimate the invariants.
As far as we know,  there  is no other algorithms with these features reported  in the literature.  For this reason,
our results  were compared against reported  estimations that were obtained  selecting the scaling  region by visual
inspection.

Regarding the  coarse-grained estimators,  the most  outstanding result was  obtained with $\CKu\E{h}$.  It  is more
accurate than previously proposed estimators and needs lesser values of $m$ to converge.

Finally,  it is  important to recall  that the implementation  of our method  only requires  the calculation  of two
correlation  integrals:  $\UCi{m}{\beta=m}\E{h}$ and  $\UCi{m}{\beta=m-2}\E{h}$,  that can  be obtained  through the
noise-assisted correlation algorithm (Alg.~\ref{alg:NCA}).  It is also important to say that the time series must be
normalized to have unitary standard deviation.  As it was reported in~\cite{Restrepo2016}, the correlation integrals
computed  using  the noise-assisted  correlation  algorithm present  oscillations  for  small  values  of $h$.  This
phenomenon makes  difficult to look for  scaling regions over small  $h$ values.  In order to  avoid this difficulty
three  strategies can  be followed:  The  first one  is to  increase  the  number  of  comparators  $\Omega$  in the
noise-assisted correlation algorithm  by increasing the data length.  This will  make coarse-grained curves smoother
and also will improve the convergence of these estimators.  The  second one is to increase $\Omega$ by making copies
of the squared distances  between delay vectors $z_{\omega}$ and adding  different noise realizations $\mu_{\omega}$
to them.  This will  not improve the convergence of  any coarse-grained estimators to  the invariant since no
new information is added,  but the ripple in the coarse-grained curves will be reduced.  The third one,  and the one
we have found as the more adequate,  is to use the algorithm proposed in~\cite{Luo2006} to calculate the logarithmic
derivatives of the  \gls{uci}.  This algorithm calculates a smoothed  version of the derivative of  a function using
the wavelet transform.

\section{\label{sec:Con}Conclusions} 

The original contributions of this document are twofold: both a new method for the estimation of invariants based on
the recently proposed U-correlation  integral,  and an algorithm for the automatic selection  of the scaling region.
The combined  use of these two  ideas allows to automatically  estimate the noise  level,  correlation dimension and
correlation entropy of a dynamical system.  This approach have been statistically tested using synthetic data coming
from low-dimensional systems with  different data lengths and noise levels.  The results  suggest that our algorithm
provides robust estimations of the correlation dimension, the correlation entropy and the noise level.
\appendix
\section[Analytic deduction for the coarse-grained \texorpdfstring{$D$}{D} estimator]
{\label{sec:A1}Analytic deduction for the coarse-grained $D$ estimator}

Let:
\begin{align}
 &a=\frac{\beta+m}{2};&&b=\frac{m-D}{2};&&c=\frac{m+2}{2};& \nonumber\\
 &y=-\frac{h^2}{4\sigma^2};&&\frac{\Di{y}}{\Di{h}}=-\frac{h}{2\sigma^2},&&\,&\label{eq:var}
\end{align}
and
\begin{equation*}
    P=\E{-1}^{m/2}\frac{\hat{\phi}}{2}\EA{-m\itau K_2}\frac{\Gf{D/2}\Gf{\E{\beta\p m}/{2}}}{\Gf{\beta/2}\Gf{m/2\p1}}
\end{equation*}
then Eq.~(\ref{eq:UCI}) can be rewritten as:
\begin{equation}
    \UCi{m}{\beta}\E{y}=Py^{m/2}\HF{a}{b}{c}{y} \label{eq:D1}  
\end{equation}

We need to find the logarithmic derivative of Eq.~(\ref{eq:D1}).  For this end we can find its first derivative as:
\begin{align}
    \dif{\Q{\UCi{m}{\beta}\E{y}}}{h}=&\dif{\Q{Py^{m/2}\HF{a}{b}{c}{y}}}{y}\dif{y}{h}\nonumber\\
    \intertext{using Eq.~15.2.1 from Ref.~\cite{Abramowitz1964}:}
    \dif{\Q{\UCi{m}{\beta}\E{y}}}{h}=&-\frac{Ph}{2\sigma^2}y^{m/2}\bigg[\frac{m}{2}\HF{a}{b}{c}{y}\nonumber\\
                                 &+\frac{ab}{c}\HF{a\p1}{b\p1}{c\p1}{y}\bigg]\nonumber
\end{align}
Then the logarithmic derivative can be found as:
\begin{align}
    \dif{\Q{\ln\UCi{m}{\beta}\E{y}}}{\ln h}=&\frac{h}{{\UCi{m}{\beta}\E{y}}}{\dif{\UCi{m}{\beta}\E{y}}{h}}\nonumber\\
    =& m+ \frac{2aby}{c}\frac{\HF{a+1}{b+1}{c+1}{y}}{\HF{a}{b}{c}{y}},\label{eq:D2}
\end{align}
and for  simplicity in  the notation:
\begin{equation*}
\logD\E{y}=\dif{\Q{\ln\UCi{m}{\beta}\E{y}}}{\ln h}.   
\end{equation*}

Using  the identity~\cite[Eq.~9.137.9]{Gradshteyn1994}:
\begin{align}
    \frac{aby}{c}\HF{a\p1}{b\p1}{c\p1}{y}=&\Frac{1\m y}\big[\E{c\m a}\HF{a\m1}{b}{c}{y}\nonumber\\
                                          &+\E{a\m c\p by}\HF{a}{b}{c}{y}\big] \label{eq:ident}
\end{align}
we can rewrite Eq.~(\ref{eq:D2}) as:
\begin{align}
    \logD\E{y}=&m+
    \frac{2}{1\m y}\Q{\E{a\m c\p by}+\E{c\m a}\frac{\HF{a\m1}{b}{c}{y}}{\HF{a}{b}{c}{y}}}
    \label{eq:D3}
\end{align}

From Eq.~(\ref{eq:UCI}) and using the definitions in (\ref{eq:var}) it is possible to demonstrate that:
\begin{equation}
    \frac{\UCi{m}{\beta-2}\E{y}}{\UCi{m}{\beta}\E{y}}=
    \frac{\beta-2}{2\E{a-1}}\frac{\HF{a-1}{b}{c}{y}}{\HF{a}{b}{c}{y}}\label{eq:D4}.
\end{equation}

Substituting Eq.~(\ref{eq:D4}) into~(\ref{eq:D3}):
\begin{align}
    \logD\E{y}=&m+\frac{2}{1\m y}\Q{\E{a\m c\p by}
    +\frac{2\E{c\m a}\E{a\m1}}{\beta\m2}\frac{\UCi{m}{\beta-2}\E{y}}{\UCi{m}{\beta}\E{y}}}\nonumber
\end{align}
Restoring the values of $\beta$, $m$ and $D$ and clearing for $D$
\begin{align}
    D=&\logD\E{y}\m\Frac{y}\Q{\logD\E{y}
    \p\E{m\p\beta\m2}\E{\frac{\UCi{m}{\beta-2}\E{y}}{\UCi{m}{\beta}\E{y}}\m1}}\nonumber
\end{align}
notice that $-1/y=4\sigma^2/h^2$ and from Eq.~(\ref{eq:delta})
\begin{equation*}
    \frac{h^2}{4\sigma^2}=\frac{\Del\E{h}}{1-\Del\E{h}}, 
\end{equation*}
then 
\begin{align}
    D=&\logD\E{h}+\frac{\Del\E{h}}{1-\Del\E{h}}\bigg[\logD\E{h}\nonumber\\
              &\p\E{m\p\beta\m2}\E{\frac{\UCi{m}{\beta-2}\E{h}}{\UCi{m}{\beta}\E{h}}\m1}\bigg].\label{eq:D5}
\end{align}

Making $\beta=m$, the coarse-grained estimator for $D$ can be defined as:
\begin{align}
        \CDu\E{h}=&\logDm\E{h}+\frac{\Del\E{h}}{1-\Del\E{h}}\bigg[\logDm\E{h}\nonumber\\
              &\p2\E{m\m1}\E{\frac{\UCi{m}{\beta=m-2}\E{h}}{\UCi{m}{\beta=m}\E{h}}\m1}\bigg].
\end{align}


\subsection[Relation with Nolte's~\glsname{etal} coarse-grained \texorpdfstring{$D$}{D} estimator]
{\label{sec:A1.1}Relation with Nolte's~\glsname{etal} coarse-grained \texorpdfstring{$D$}{D} estimator:}

It  was  proved  in~\cite{Restrepo2016}  that  the  Gaussian  correlation  integral  is  a  particular  case  of the
U-correlation  integral  that  arises  when  $\beta$  is  constant  and  equal  to  $2$.  Let  $\beta=2$,  then from
Eq.~(\ref{eq:D4}) it can be proved that:
\begin{equation*}
    \frac{\UCi{m}{\beta-2}\E{h}}{\UCi{m}{\beta}\E{h}}=0,
\end{equation*}
for all $m$ values. Thus, Eq.~(\ref{eq:D5}) can be rewritten as:
\begin{align}
  D=&\logDmb{\beta=2}{m}\E{h}-\frac{\Del\E{h}}{1-\Del\E{h}}\E{m-\logDmb{\beta=2}{m}\E{h}}. \label{eq:D6}
\end{align}

Finally,  given  that the  Gaussian correlation  integral is  a particular  case of  the U-correlation  integral for
$\beta=2$, i.e.,  ${\UCi{m}{\beta=2}\E{h}=T_{m}\E{h}}$, we can write:
\begin{equation*}
  \logDmb{\beta=2}{m}\E{h}=\dif{\Q{\ln T_{m}\E{h}}}{\ln h}.
\end{equation*}

Then,  Eq.~(\ref{eq:D6}) is  the same coarse-grained  estimator for the  correlation dimension  that was  derived by
Nolte~\glsname{etal} in~\cite{Nolte2001}.
\section[Analytic deduction for the coarse-grained \texorpdfstring{$K_2$}{K2} estimator]
{\label{sec:A2}Analytic deduction for the coarse-grained $K_2$ estimator}

The analytic deduction of the coarse-grained estimator for \gls{k2} begins by defining the quantity:
\begin{equation*}
    M=\frac{\hat{\phi}}{2}\E{2\sigma^2}^{D}\Gf{D/2}.
\end{equation*}
Substituting this equation and Eq.~(\ref{eq:var}) into Eq.~(\ref{eq:UCI}) we can write:
\begin{equation*}
    \UCi{m}{\beta}\E{y}=M\frac{\Gf{a}}{\Gf{\beta/2}\Gf{c}}\EA{-m\itau K_2}\E{-y}^{m/2}\HF{a}{b}{c}{y} \label{eq:K1}  
\end{equation*}
Then the ratio  $\UCi{m+2}{\beta+2}\E{y}/\UCi{m}{\beta}\E{y}$ can be written as:
\begin{equation*}
    \frac{\UCi{m+2}{\beta+2}\E{y}}{\UCi{m}{\beta}\E{y}}=\frac{2a\E{a\p1}}{\beta c}\EA{-2\itau
    K_2}\E{-y}\frac{\HF{a\p2}{b\p1}{c\p1}{y}}{\HF{a}{b}{c}{y}}.
\end{equation*}

Making $\alpha=a+1$: 
\begin{equation}
    \frac{\UCi{m+2}{\beta+2}\E{y}}{\UCi{m}{\beta}\E{y}}=\frac{2\E{\alpha\m1}\alpha}{\beta c}\EA{-2\itau
    K_2}\E{-y}\frac{\HF{\alpha\p1}{b\p1}{c\p1}{y}}{\HF{\alpha\m1}{b}{c}{y}}. \label{eq:K2}
\end{equation}

The identity in Eq.~(\ref{eq:ident}) can be rewritten as:
\begin{align}
    \frac{\alpha \E{-y}}{c}\frac{\HF{\alpha\p1}{b\p1}{c\p1}{y}}{\HF{\alpha\m1}{b}{c}{y}}=&\frac{-1}{b\E{1\m
    y}}\bigg[\E{c\m \alpha}\nonumber\\
                                                                                     &+\E{\alpha\m c\p
    by}\frac{\HF{\alpha}{b}{c}{y}}{\HF{\alpha\m1}{b}{c}{y}}\bigg]. \label{eq:ident2}
\end{align}

Substituting Eq.~(\ref{eq:ident2}) into Eq.~(\ref{eq:K2}) and restituting $a$:
\begin{align}
    \frac{\UCi{m+2}{\beta+2}\E{y}}{\UCi{m}{\beta}\E{y}}=&\frac{-2a}{\beta b\E{1-y}}\EA{-2\itau
    K_2}\bigg[\E{c\m a\m1}+\nonumber\\
    &+\E{a\p1\m c\p by}\frac{\HF{a\p1}{b}{c}{y}}{\HF{a}{b}{c}{y}}\bigg].\label{eq:K3}
\end{align}

Using the identity~\cite[Eq.~9.137.12]{Gradshteyn1994}: 
\begin{align}
    \frac{by}{c}\frac{\HF{a+1}{b+1}{c+1}{y}}{\HF{a}{b}{c}{y}}+1=\frac{\HF{a+1}{b}{c}{y}}{\HF{a}{b}{c}{y}}.
\end{align}

Equation~(\ref{eq:K3}) can be written as:
\begin{align}
    \frac{\UCi{m+2}{\beta+2}\E{y}}{\UCi{m}{\beta}\E{y}}=&\frac{\EA{-2K_2\itau}}{2b\beta\E{y-1}}\bigg[4a\E{c\m a\m1}+\nonumber\\
                                                        &+2\E{a\p1\m c\p
    by}\E{\frac{2aby}{c}\frac{\HF{a\p1}{b\p1}{c\p1}{y}}{\HF{a}{b}{c}{y}}+2a}\bigg].
\end{align}

Using Eq.~(\ref{eq:D3})
\begin{align}
    \frac{\UCi{m+2}{\beta+2}\E{y}}{\UCi{m}{\beta}\E{y}}=&\frac{\EA{-2K_2\itau}}{2b\beta\E{y-1}}\bigg[4a\E{c\m a\m1}+\nonumber\\
                                                        &+2\E{a\p1\m c\p by}\E{\logD\E{y}-m+2a}\bigg].
\end{align}

Replacing the values of $a$, $c$ and $b$: 
\begin{align}
    \frac{\UCi{m+2}{\beta+2}\E{y}}{\UCi{m}{\beta}\E{y}}=&\frac{\EA{-2K_2\itau}}{2b\beta\E{y\m1}}\Q{\m\beta\E{m\p\beta}\p\E{\beta\p
    2by}\E{\logD\E{y}\p \beta}}\nonumber\\
                                                    =&\frac{\EA{-2K_2\itau}}{2b\E{y\m1}}\bigg[ \logD\E{y}\m m\p
                                                     {2by}\E{{\frac{\logD\E{y}}{\beta}\p1}}\bigg]\nonumber\\
                                                    =&\frac{\EA{-2K_2\itau}}{\E{y\m1}}\Q{\frac{\logD\E{y}\m m}{m\m D}
                                                     +y\E{{\frac{\logD\E{y}}{\beta}\p1}}}.
\end{align}
It can be shown that $\Del\E{h}=1/\E{1-y}$ and  $1-\Del\E{h}=y/\E{y-1}$ so:
\begin{align}
    \frac{\UCi{m+2}{\beta+2}\E{y}}{\UCi{m}{\beta}\E{y}}=&\EA{-2K_2\itau}\Bigg[\Del\E{h}\E{\frac{m\m\logD\E{y}}{m\m D}}+\nonumber\\
+&\E{1\m \Del\E{h}}\E{{\frac{\logD\E{y}}{\beta}\p1}}\Bigg].\nonumber
\end{align}
Taking natural logarithm, subtracting the term $\ln\E{D/\beta+1}$ on both sides and dividing by $2\itau$:
\begin{align}
    K_2-\Frac{2\itau}\ln\E{\frac{D}{\beta}+1}=&\Frac{2\itau}\Bigg\{\ln\Bigg[\Del\E{h}\E{\frac{m\m\logD\E{y}}{m\m D}}+\nonumber\\
    +&\E{1\m\Del\E{h}}\E{{\frac{\logD\E{y}}{\beta}\p1}}\Bigg]\nonumber\\
-&\ln\Q{\frac{\UCi{m+2}{\beta+2}\E{y}}{\UCi{m}{\beta}\E{y}}}-\ln\E{\frac{D}{\beta+1}}\Bigg\}.\label{eq:K4}
\end{align}

Let call the left-hand side of Eq.~(\ref{eq:K4}) as $\CKu\E{h}$:
\begin{align}
    \CKu\E{h}=&\Frac{2\itau}\Bigg\{\ln\Bigg[\Del\E{h}\E{\frac{m\m\logD\E{h}}{m\m D}}+\nonumber\\
    +&\E{1\m\Del\E{h}}\E{{\frac{\logD\E{y}}{\beta}\p1}}\Bigg]-\nonumber\\
-&\ln\Q{\frac{\UCi{m+2}{\beta+2}\E{h}}{\UCi{m}{\beta}\E{h}}}-\ln\E{\frac{D}{\beta}+1}\Bigg\}.
\end{align}

Note that this estimator  still depends on the correlation dimension $D$.  However,  this  dependency can be avoided
using $\CDu\E{h}$ as estimator for $D$.  Making $\beta=m$ we can define the coarse-grained entropy estimator as:
\begin{align}
    \CKu\E{h}=&\Frac{2\itau}\Bigg\{\ln\Bigg[\Del\E{h}\E{\frac{m\m\logDm\E{h}}{m\m \CDu\E{h}}}+\nonumber\\
    +&\E{1\m\Del\E{h}}\E{{\frac{\logDm\E{y}}{m}\p1}}\Bigg]-\nonumber\\
-&\ln\Q{\frac{\UCi{m+2}{\beta=m+2}\E{h}}{\UCi{m}{\beta=m}\E{h}}}-\ln\E{\frac{\CDu\E{h}}{m}+1}\Bigg\}.
\end{align}

\bibliographystyle{spmpsci}
\bibliography{NonLinDynamics}
\end{document}